\documentclass[preprint,showpacs,preprintnumbers,amsmath,amssymb,superscriptaddress,nofootinbib]{revtex4}
\usepackage{graphicx,color}
\usepackage{amsmath,amssymb}
\usepackage{url}
\usepackage{epstopdf}
\newcommand{\hs}{\hspace*{0.5cm}}

\newcommand{\be}{\begin{equation}}
\newcommand{\ee}{\end{equation}}
\newcommand{\bea}{\begin{eqnarray}}
\newcommand{\eea}{\end{eqnarray}}
\newcommand{\nn}{\nonumber}
\newcommand{\crn}{\nonumber \\}

\newcommand{\al}{\alpha}
\newcommand{\la}{\lambda}
\newcommand{\bet}{\beta}
\newcommand{\ga}{\gamma}
\newcommand{\va}{\varphi}

\newcommand{\pa}{\partial}
\newcommand{\fr}{\frac}

\newcommand{\bc}{\begin{center}}
\newcommand{\ec}{\end{center}}
\newcommand{\Ga}{\Gamma}

\newcommand{\si}{\sigma}

\newcommand {\ba}{\begin{array}}
\newcommand {\ea}{\end{array}}
\newcommand{\ben}{\begin{enumerate}}
\newcommand{\een}{\end{enumerate}}

\usepackage{bm}
\usepackage{dcolumn}

\begin{document}

\title{Probing neutrino and Higgs sectors in  $\mbox{SU(2)}_1 \times
\mbox{SU(2)}_2 \times \mbox{U(1)}_Y $ model with lepton-flavor non-universality}
\author{L. T. Hue}
\email{lthue@iop.vast.vn}
\affiliation{Institute of Research and Development, Duy Tan University,
   Da Nang City, Vietnam}
\affiliation{Institute of Physics,   Vietnam Academy of Science and Technology, 10 Dao Tan, Ba
Dinh, Hanoi, Vietnam }
\author{A. B. Arbuzov}\email{arbuzov@theor.jinr.ru}
\address{Bogoliubov Laboratory for Theoretical Physics, Joint Institute for Nuclear Researches, Dubna,  Russia}
\author{N.  T. K.  Ngan}\email{ntkngan@ctu.edu.vn}
\affiliation{Department of Physics, Cantho University,
3/2 Street, Ninh Kieu, Cantho, Vietnam}
\affiliation {Graduate University of Science and Technology, Vietnam Academy of Science and Technology, 18 Hoang Quoc Viet, Cau Giay, Hanoi, Vietnam}
\author{H.  N.  Long}\email{hoangngoclong@tdt.edu.vn}
\address{Theoretical Particle Physics and Cosmology Research Group, Ton Duc Thang University, Ho Chi Minh City, Vietnam}
\address{Faculty of Applied Sciences,
 Ton Duc Thang University, Ho Chi Minh City, Vietnam}

\begin{abstract}
The  neutrino and Higgs sectors in the $\mbox{SU(2)}_1 \times \mbox{SU(2)}_2 \times \mbox{U(1)}_Y $ model with lepton-flavor non-universality are discussed. We show that active neutrinos can get Majorana masses from radiative corrections, after adding only new singly charged Higgs bosons. The mechanism for generation of neutrino masses is the same as in the Zee models. This also gives a hint to solving the dark matter problem based on similar ways discussed recently in many radiative neutrino mass models with dark matter. Except the active neutrinos, the appearance of singly charged Higgs bosons and dark matter does not affect significantly the physical spectrum of all particles in the original model.
 We indicate this point by investigating the Higgs sector in both cases before and after singly charged scalars are added into it. Many interesting properties of physical Higgs bosons, which were not shown previously, are explored. In particular, the mass matrices of charged and CP-odd Higgs fields are proportional to the coefficient of triple Higgs coupling $\mu$. The mass eigenstates and eigenvalues in the CP-even Higgs sector are also presented. All couplings of the SM-like Higgs boson to normal fermions and gauge bosons are different from the SM predictions by a factor $c_h$, which must satisfy the recent global fit of experimental data, namely $0.995<|c_h|<1$.
  We have analyzed a more general diagonalization of gauge boson mass matrices, then we show that the ratio
   of the tangents of the  $W-W'$ and   $Z-Z'$  mixing angles is exactly the cosine of the Weinberg angle, implying that number of parameters is reduced by 1. Signals of new physics from decays of new heavy fermions and Higgs bosons at LHC and constraints of their masses are also discussed.


\end{abstract}


\maketitle
\section{Introduction}
\label{intro}
One of the most important purposes of the LHC is to search for  manifestations of new physics (NP).
It seems that some clues have appeared with massive neutrinos and recent observations of lepton-flavor non-universality (LNU).
Recall that the lepton family replication is assumed in the Standard Model (SM). Therefore, the lepton-flavor is universal in the latter. For the recent two decades,
neutrino and Higgs physics are hot topics in Particle Physics.
With increasing luminosity and beam energy, the LHC becomes a powerful tool for searching for NP.
With larger masses, the third generation seems to be more interesting, in the sense of the
sensitivity to NP. Nowadays, there are two kinds of anomalies in the semileptonic $B$ meson
decays which are captivating for the LNU. The first one is the class of the following ratios
of branching fractions:
 \bea
R_{D^*} & = & \fr{\Ga(\bar{B} \rightarrow D^*\, \tau\, \tilde{\nu} )}{\Ga(\bar{B} \rightarrow
D^*\, l\, \tilde{\nu} )} = 0.310 \pm 0.015 \pm 0.008\, , \crn
R_D & = & \fr{\Ga(\bar{B} \rightarrow D\, \tau\, \tilde{\nu} )}{\Ga(\bar{B} \rightarrow D\, l\, \tilde{\nu} )} = 0.403 \pm 0.040\pm 0.024
\, , \hs l = e, \mu,
\label{g221e1}
\eea

which show $3.5 \, \si$ deviations from the corresponding SM predictions \cite{sak14},
\be
R_{D^*} = 0.252 \pm 0.004\, , \hs R_D=0.305 \pm 0.012.
\label{g221e2}
\ee
The above results provide hints for violation of the lepton flavor universality (LFU).

The second kind of anomalies~\footnote{ The SM value for $R_K$  has been first obtained in Ref.\cite{hiller}}
is the interesting LNU ratio reported recently by LHCb \cite{lhcb1}, namely,
\be
R_K = \fr{\Ga(B \rightarrow K\, \mu^+\, \mu^- )}{\Ga(B \rightarrow K\, e^+\, e^- ))}
= 0.745^{+0.090}_{-0.074} \pm 0.036\, ,
\label{g221e3}
\ee
which has $2.6 \, \si$ deviation from the SM value $R_K = 1.0003 \pm 0.0001$
in the dilepton mass squared bin $(1 \leq q^2 \leq 6) \, \textrm{GeV}^2$.

 From the physical point of view, the mass of a particle plays a quite important role
in its characteristic properties. To justify this, let us mention some well-known examples.
The first is that the proton and neutron have a tiny mass difference (940  vs 938) MeV,
but the proton is long-lived while the neutron is unstable with its mean lifetime
of just under 15 min ($881.5 \pm 1.5$ s). The second example is the situation with the
electron and the muon. Both particles are leptons with just a mass difference
(0.511 vs 105.6) MeV, the electron is stable while the muon is unstable with the mean lifetime of 2.2 $\mu$s.
So one may expect that the third generation of quarks and leptons where particles are heavier,
has to be different from the first two ones. Within this context, the above data showing
the LNU look quite understandable. In other words it is quite natural to expect that the third
fermion generation is more strongly coupled to some New Physics than the first two ones.
Recently the $R_D$ and $R_{D^*}$ were subjects of intensive studies mostly in scalar
leptoquark models\cite{lqmodel,lqmodelm}.

One of the beyond the SM models satisfying the recent experimental data of LNU
is the model based on the $\mbox{SU(2)}_1 \times \mbox{SU(2)}_2 \times \mbox{U(1)}_Y $
(G221) gauge group~\cite{g221} (more kinds of G221 models can be found in Ref.~\cite{gg221}).
In Ref.~\cite{g221} the authors have mainly concentrated on explanation of LNU
in the lepton sector. But at present, any theoretical model in particle physics
has to deal with neutrino masses, the baryon asymmetry of the universe (BAU) and
the dark matter (DM).

The aim of this work is to study further details in the gauge, Higgs and neutrino sectors
of the model presented in Ref.~\cite{g221}.  We will show that the problems of the active
neutrino mass and DM in this model can be solved without any changes of results of allowed
parameter regions satisfying all constraints of the flavor physics, tau decays, electroweak
precision data, and recent anomalies in $B$ decays, which were indicated in Ref.~\cite{g221}.
In particular, the active neutrinos get Majorana masses from radiative corrections, where
new lepton-number violating interactions have to be introduced. The simplest way is the Zee
method~\cite{Zee}, where a pair of singly charged scalars transformed as singlets under both
the $SU(2)$ gauge groups is introduced. Like in the Zee models, where a second $SU(2)_L$ Higgs
doublet is necessary for creating a nonzero triple coupling of two Higgs doublets and a singly
charged Higgs singlet, the  $SU(2)_1$ Higgs doublet $\phi'$ in this G221 model plays the role
of the second $SU(2)_L$ Higgs doublet. Hence, no new breaking scales need to appear, implying
that there are no new mass terms contributing to the fermion and gauge boson sectors.
This explains why all results investigated in Ref.~\cite{g221} are unchanged, therefore we can
use them to study the coupling properties of the Higgs and gauge bosons with fermions.
In addition, it suggests that the ways of generating active neutrino masses in many recent
radiative neutrino mass models can be applied to the G221 model. Many of these  models have DM
candidates that are neutral fermion singlets and have odd charges under
\textbf{a} new $Z_2$ symmetry.
To avoid complicate Higgs sectors, where just new charged Higgs bosons are included,
we will not pay much attention to models solving the DM problems in this work.
We will discuss in detail the mechanism of generating neutrino masses from the Zee mechanism,
and the  Higgs potential with the appearance of two singly charged Higgs singlets.
In the gauge boson sector, we  will apply general method to diagonalize neutral and charged gauge  boson sectors,
and from this we get a consequence that the tangents of the  mixing angles in two sectors are proportional.
This will reduce the number of the model parameters by 1. In the Higgs sector, the physical Higgs spectrum
is presented. Then the SM-like Higgs boson and its couplings to other SM-like particles are identified
and compared with the SM predictions. A comparison between properties of the Higgs spectrum in the G221 model
and the minimal supersymmetric model (MSSM) and two Higgs doublet models (THDM) will also be  discussed in this work.
Based on these properties and the constraints of parameters given in~\cite{g221}, we will discuss
the bounds of new Higgs boson masses as well as promoting decay channels of the Higgs bosons
and fermions that can be searched for at modern colliders such as the LHC.

This paper is organized as follows.  In section~\ref{brief}, after a brief review of the model,
we present a more careful consideration of charged lepton masses and the Zee method for generation
of neutrino masses. In the subsection~\ref{neutrallepton} we suggest two possibilities of appearance
of DM candidates in the G221 model. The first is based on a radiative neutrino mass model introduced
previously. This way will not change the results of parameter constraints in Ref.~\cite{g221}.
The second way is different, because a new scalar $SU(2)_2$ triplet is included.
It contains a new neutral component with nonzero vacuum expectation value (VEV), leading to a new mass
scale contributing to gauge boson masses. But this model may predict some active neutrinos playing
the role of DM candidates. A more careful diagonalization of squared mass matrices and mixing parameters
of gauge bosons is presented in section~\ref{gaugeboson}. In this section, the relation between
the tangents of the  $W-W'$ and   $Z-Z'$  mixing angles, is derived. Then a validation of  $\rho$ parameter under recent
 experimental constraint will be shown at TeV scale of $SU(2)_1$ breaking scale.
Section~\ref{current} is devoted to charged and neutral currents in the model. Here we notice
their difference from the SM ones. For the further discussion of the NP searches at colliders,
the couplings of $Z$ and $W$ gauge bosons with fermions are given.
In section~\ref{Higgssector} a detailed analysis of the Higgs sector is presented.
This section covers both versions of the Higgs sector content without and with the mentioned charged scalars.
And the SM-like Higgs boson is identified. Interesting properties of singly charged Higgs bosons
are also discussed. In section~\ref{pheno}, we review briefly the allowed regions of parameters
given in~\cite{g221}, which resulted from a specific numerical illustration in the limit of two  vector-like fermion generations and simple textures of Yukawa couplings.  Following the searches for new heavy particles at the LHC, we use these
allowed regions to investigate lower bounds of masses and promoting  decay channels of new fermions
and Higgs bosons predicted by this model.
Conclusions are given in the last section~\ref{conclusion}.
\section{Brief review of the model}
\label{brief}
The model is based on the gauge group $\mbox{SU(2)}_1 \times \mbox{SU(2)}_2 \times \mbox{U(1)}_Y $
with the following gauge couplings, fields and generators~\cite{g221}:
\bea
\mbox{SU(2)}_1 &:& g_1 \, , W_i^1\, , T_i^1\, , \crn
\mbox{SU(2)}_2 &:& g_2 \, ,   W_i^2\, ,  T_i^2\, ,\label{g221e5}\\
\mbox{U(1)}_Y &:& g' \, ,  B\, ,  Y \, ,
\nn
\eea
where $i= 1, 2, 3$ is the $\mbox{SU(2)}$ index.  All the chiral fermions transform as
\allowdisplaybreaks
\bea
q_L & \sim & \left(\textbf{3,1,2},\fr 1 6\right)\, ,\, \ell_L \sim \left(\textbf{1, 1, 2}, -\fr 1 2\right) \, ,\crn
u_R & \sim & \left( \textbf{3, 1, 1}, \fr 2 3\right)\, ,\,  e_R \sim \left(\textbf{1, 1, 1}, -1\right)\, ,\;
d_R  \sim  \left(\textbf{3, 1, 1}, -\fr 1 3\right) \, ,
\label{g221e6}
\eea
where the numbers in brackets refer to $\mbox{SU(3)}_C$, $\mbox{SU(2)}_1$, $\mbox{SU(2)}_2$,
and the hypercharge. The electric charge operator is determined in the form
\be
Q = (T_3^1 + T_3^2) + Y.
\label{g221e7}
\ee
For the subgroup $\mbox{SU(2)}_1 $ there are $n_{VL}$  generations of \emph{vector-like} fermions
which are transformed as its doublets, while they are singlets for the $\mbox{SU(2)}_2$,
\be Q_{L,R} \equiv
\left(%
\begin{array}{c}
U \\
D\
\end{array}\,
\right)_{L,R} \sim \left( \textbf{3, 2, 1}, \fr 1 6 \right)\,  ; \,  L_{L,R} \equiv
\left(%
\begin{array}{c}
N \\
E\
\end{array}\,
\right)_{L,R} \sim \left( \textbf{1, 2, 1}, -\fr 1 2 \right) \, .
\label{g221e8}
\ee
The vector-like fermion generation number is greater than one in order to explain successfully the LNU,  and it was fixed by $n_{VL}=2$ for simplicity in numerical illustration \cite{g221}.

The Higgs sector consists of two doublets $\phi$ and $\phi'$ and one self-dual bidoublet $\Phi$
(i.e., $\Phi = \si_2 \Phi^* \si_2$ where $\si_2$ is the usual Pauli matrix)
\be
\phi \sim \left(\textbf{1,1,2}, \fr 1 2 \right)\, , \phi' \sim \left(\textbf{1,2,1}, \fr 1 2 \right)\, ,
\Phi \sim (\textbf{1, 2}, \tilde{{\bf 2}},0) \, ,
\label{g221e9}
\ee
with components as
\be \phi = \left(%
\begin{array}{c}
\va^+ \\
\va^0\
\end{array}\,
\right) \, , \phi' = \left(%
\begin{array}{c}
\va'^+ \\
\va'^0\
\end{array}\,
\right)\, , \Phi = \fr{1}{\sqrt{2}} \left(%
\begin{array}{cc}
\Phi^0 & \Phi^+ \\
-\Phi^-
& \tilde{\Phi}^0\\
\end{array}\,
\right)\, ,
\label{g221e10}
\ee
with $\tilde{\Phi}^0 = (\Phi^0)^*$.
The scalar fields develop VEVs
\be \langle \phi \rangle = \fr{1}{\sqrt{2}} \left(%
\begin{array}{c}
0\\
v_\phi\
\end{array}\,
\right) \, , \langle \phi' \rangle =\fr{1}{\sqrt{2}} \left(%
\begin{array}{c}
0 \\
v_{\phi'}\
\end{array}\,
\right)\, , \langle \Phi \rangle = \fr{1}{2} \left(%
\begin{array}{cc}
u & 0\\
0
& u\\
\end{array}\,
\right)\, .
\label{g221e11}
\ee
The spontaneous symmetry breaking (SSB) of the model follows the pattern
\be
\mbox{SU(2)}_1 \times \mbox{SU(2)}_2 \times \mbox{U(1)}_Y  \stackrel{u}\longrightarrow
\mbox{SU(2)}_L \times \mbox{U(1)}_Y
\stackrel{ v_\phi,\,  v_{\phi'} }\longrightarrow   U(1)_Q\, .
\label{g221e11t}
\ee
The main phenomenology of the model concerned $B$-decay anomalies and the lepton-flavor non-universality has
been presented in~\cite{g221}. However, the current physical model has to satisfy Higgs and neutrino physics as well as DM candidate.

With the above breaking chain, the VEVs are assumed to satisfy the relation
\be
u \gg v_\phi, v_{\phi'}\, .
\label{g221e11tn}
\ee
Yukawa Lagrangian,  fermion mass  matrices, and  diagonalization steps to  construct physical states and masses of fermions were presented in  detail  in \cite{g221}. Hence, we will summarize here only important results and focus on  new features of generating  active neutrino masses from loop corrections.
\subsection{Charged fermion masses}
\label{chargedgauge}
The chiral fermions couple to the SM Higgs-like $\phi$ doublet
\be
 -{\cal L}_\phi = {\bar q_L}\,  y_d \phi\,  d_R + {\bar q_L}\,  y_u \, \tilde{\phi}\,
u_R + {\bar \ell_L}\,  y_\ell \, \phi \, e_R\, + \mathrm{H.c.},
 \label{g221e57}
 \ee
where $\tilde{\phi} \equiv i\si_2 \phi^*$. The matrices $y_d, y_u, y_\ell$ are $3 \times 3$ matrices.
The vector-like fermions \emph{can have gauge-invariant Dirac mass terms}
\be
 -{\cal L}_M = {\bar Q_L}\, M_Q \,  Q_R + {\bar L_L}\,  M_L  L_R  + \mathrm{H.c.}
 \label{g221e58}
\ee
Other contributions are
\bea
 -{\cal L}_\Phi &=& {\bar Q_R}\,  \la^\dag_q \, \Phi\,   q_L + {\bar L_R}\, \la^\dag_\ell \,\Phi\,
\ell_L \, + \mathrm{H.c.} , \label{g221e60}\\
-{\cal L}_{\phi '} &=& {\bar Q_L}\,  {\tilde y}_d \phi'\,  d_R + {\bar Q_L}\,  {\tilde y}_u \,
\tilde{\phi'}\,  u_R + {\bar L_L}\,  {\tilde y}_\ell \, \phi' \, e_R\, +\mathrm{ H.c.} ,
 \label{g221e59}
 \eea
where $\la^\dagger_{q,\, \ell}$ and ${\tilde y}_{u, \, d,\,\ell}$ are $n_{VL} \times 3$ Yukawa matrices.
After the SSB, \emph{the above couplings will induce mixing between the vector-like and the SM chiral}
fermions. This is crucial for the phenomenology of the model.

 For the sake of simplicity one can assume a softly broken discrete ${\cal Z}_2$ symmetry under which only $\phi '$ is odd, making unnecessary Yukawa couplings vanish, i.e., ${\tilde y}_{u, \, d,\,\ell} \simeq 0$ \cite{g221}. There is another charge assignment  that also forbids Lagrangian in (\ref{g221e59}), while keeps $\phi'$ even:  only $Q_L$ and $L_L$  are odd. This  is necessary for generating active neutrino masses by the Zee method considered in this work.

We combine  the  chiral and vector-like fermions as
\bea
{\cal U}^I_{L,R} & \equiv & (u^i_{L,R}, U^k_{L,R})^T\, , \,
{\cal D}^I_{L,R} \equiv (d^i_{L,R}, D^k_{L,R})^T\, , {\cal E}^I_{L,R}  \equiv  (e^i_{L,R}, E^k_{L,R})^T,
\label{g221e61}
 \eea
where $i=1, 2, 3$, $k=1, \cdots, n_{VL} $ and $I = 1, \cdots, 3+n_{VL}$.
After the SSB, the fermion mass Lagrangian has the form
\be
- {\cal L}_{mass}^f = {\bar {\cal U}_L} {\cal M_U} {\cal U}_R + {\bar{\cal D}_L} {\cal M_D} {\cal D}_R
+ {\bar{\cal E}_L} {\cal M_E} {\cal E}_R + \mathrm{H.c.}
\label{g221e62}
\ee
Here, all above mass matrices are $(3+n_{VL}) \times (3+n_{VL})$ and have the form
\be
{\cal M_U} =  \left(%
\begin{array}{cc}
\fr{1}{\sqrt{2}}y_u v_\phi  & \fr 1 2 \la_q u \\
\fr{1}{\sqrt{2}}{\tilde y}_u v_{\phi '}
& M_Q\\
\end{array}\,
\right)\, , \,  {\cal M_D} =  \left(%
\begin{array}{cc}
\fr{1}{\sqrt{2}}y_d v_\phi  & \fr 1 2 \la_q u \\
\fr{1}{\sqrt{2}}{\tilde y}_d v_{\phi '}
& M_Q\\
\end{array}\,
\right), \;
 {\cal M_E} =  \left(%
\begin{array}{cc}
\fr{1}{\sqrt{2}}y_{\ell} v_\phi  & \fr 1 2 \la_\ell u \\
\fr{1}{\sqrt{2}}{\tilde y}_{\ell} v_{\phi '}
& M_L\\
\end{array}\,
\right).
\label{g221e63}
 \ee
In the limit $\epsilon=v/u\ll 1$, these matrices are  blocked-diagonalized perturbatively via two steps. After that, the SM  parts are separated from the total. The transformations of fermion states are: ${\cal U}_L\rightarrow V^{\dagger}_QV^{\dagger}_u{\cal U}_L$,  ${\cal D}_L\rightarrow V^{\dagger}_QV^{\dagger}_d{\cal D}_L$,  ${\cal E}_L\rightarrow V^{\dagger}_LV^{\dagger}_e{\cal E}_L$, ${\cal U}_R\rightarrow W^{\dagger}_u{\cal U}_R$, ${\cal D}_R\rightarrow W^{\dagger}_d{\cal D}_R$, and ${\cal E}_R\rightarrow W^{\dagger}_e{\cal E}_R$, where $V_{F}$ ($F=Q,L$), $V_{f}$ ($f=e,u,d$), and $W_f$ are $(3+n_{V_L})\times (3+n_{V_L})$ unitary matrices  \cite{g221}. At the first step where  $v=0$, every ${\cal M_F}$ ($ {\cal  F = U, D, E}$) is diagonalized by an exact $V_{F}$ depending on $u,\, M_{F}$ and $\lambda_{\ell,q}$.  At the second step, transformations $V_{f}$ and $W_{f}$ are expanded in terms of power series of $\epsilon$, $V_f= 1+i \epsilon^2 H^f_V+ ...$ and $W_f= 1+i\epsilon H^f_W+1/2 (i\epsilon H^f_W)^2+...$.  They were listed precisely in \cite{g221}. After the two  steps, all original mass matrices in (\ref{g221e63}) will be transformed into block-diagonal forms $\hat{{\cal M}}_{ {\cal F}}=V_fV_{ F}{\cal M}_{ {\cal F}}W^{\dagger}_f$. One of the blocks in every $\hat{{\cal M}}_{ {\cal F}}$ is  identified with a SM fermion block, which is diagonalized by  $3\times3$ unitary transformations: $f_L\rightarrow S^{\dagger}_f f_L$ and $f_R\rightarrow U^{\dagger}_f f_R$.  Only the CKM matrix, $V_{\mathrm{CKM}}=S_uS^{\dagger}_d$,  appears in the gauge couplings \cite{g221}. We can fix $S_e=U_e=I_3$.

For studying Higgs boson phenomenology satisfying  the allowed regions of parameters given in \cite{g221}, which resulted from a specific  assumption of two new lepton families and  textures  of Yukawa couplings $\lambda_{q,\ell}$,  we will present more detailed masses and eigenstates of charged leptons. The quark sector can be derived similarly. In the flavor basis $\mathcal{E}$
of charged leptons,  the mass matrix ${\cal M_E}$ in (\ref{g221e63}) is  $5\times 5$.
Following  Ref.~\cite{g221}, a simple texture of  $\lambda_\ell$  is  chosen as
\be
\la_{\ell}= 2 c_{\beta'} \left(
                                \begin{array}{cc}
                                  \widetilde{M}_{L_1} & 0 \\
                                  0 & \widetilde{M}_{L_2}\Delta_{\mu} \\
                                  0 & \widetilde{M}_{L_2}\Delta_{\tau} \\
                                \end{array}
                              \right),
\label{laell1}
\ee
where new parameters $ \Delta_{\mu}$ and $\Delta_{\tau}$ will be considered as free parameters;
while $ \widetilde{M}_{L_1}, \widetilde{M}_{L_2}$  are "reduced" masses of new charged leptons, $m_{E_k}\simeq u\widetilde{M}_{L_k}$ \cite{g221},
\bea
\widetilde{M}_L= \mathrm{diag} \left( \widetilde{M}_{L_1},\;
\widetilde{M}_{L_2}\right) = \sqrt{\frac{M^{\dagger}_L M_L}{u^2}+ \frac{\lambda^{\dagger}_{\ell}\lambda_{\ell}}{4}}.
\label{titleML}
\eea
We recall here important properties of  charged lepton
	parameters used  in  constructing radiative active neutrino masses. According to~\cite{g221}, physical masses $(m_{e_i},  m_{E_k})$ related to  $\mathcal{M}_{\mathcal{E}}$ in~(\ref{g221e63}) by $V_eV_{L} \mathcal{M}_{\mathcal{E}}W_{e}^{\dagger}= \mathrm{diag}(m_{e_i},\; m_{E_k})$,
the mass bases of left-and right-handed leptons
$\mathcal{E}^{(\mathrm{d})I}_{L,R}\equiv (e^{(\mathrm{d})i}_{L,R},\; E^{(\mathrm{d})k}_{L,R})^T$ are defined as
$ \mathcal{E}_{L}=V^{\dagger}_LV^{\dagger}_e \mathcal{E}^{(\mathrm{d})}_{L}$ and
$ \mathcal{E}_{R}=W^{\dagger}_e \mathcal{E}^{(\mathrm{d})}_{R}$.
The product $V_eV_{L}$ can be found from the relation
$V_eV_{L} \mathcal{M}_{\mathcal{E}}\mathcal{M}_{\mathcal{E}}^{\dagger} \left(V_eV_{L}\right)^{\dagger}=\mathrm{diag}(m^2_{e_i},\; m^2_{E_k})$. Non-diagonal elements of  $V_L$ may  be large because  those of $\mathcal{M}_{\mathcal{E}}$  are at the $SU(2)_1$ scale.
In contrast, those of  $V_e$ and $W_e$ are at least one order of  $\frac{v_{\phi}}{u}$,
because these elements of $V_{L} \mathcal{M}_{\mathcal{E}}$ are  order of the electroweak scale. Hence, $V_e$ and $W_e$ are nearly identical  when $u \gg v_{\phi}$.
They only play the role of generating light  charged lepton masses of  $e,\;\mu$, and $\tau$.
Hence, in many cases we can use the approximations $\mathcal{E}_{L}=V^{\dagger}_L\mathcal{E}^{(\mathrm{d})}_{L}$
and $ \mathcal{E}^{I}_{R}= \mathcal{E}^{(\mathrm{d})I}_{R}$. We can see that the $V_L$ is exactly
the mixing matrix of neutrinos if they are all considered as the pure Dirac particles.
Formula of $V_L$ is written in the block form, namely~\cite{g221}
\be V_L=\left(
        \begin{array}{c|c}
           V^{11}_L=\sqrt{I_3-\frac{1}{4}\lambda_{\ell}\widetilde{M}_L^{-2}\lambda^{\dagger}_{\ell}} \;\; &
           V^{12}_L=-\frac{u}{2}V^{11}_L\lambda_{\ell}M^{-1}_L \\
           \hline
           V^{21}_L=\frac{1}{2} \widetilde{M}_L^{-1}\lambda^{\dagger}_{\ell}   &
           V^{22}_L=\frac{1}{u} \widetilde{M}_L^{-1}M_L^{\dagger}\\
        \end{array}
        \right), \label{VL}
\ee
where analytic expression of $V_L^{ij}$, with $i,j=1,2$, corresponding to $\lambda_\ell$
in Eq.~(\ref{laell1}) are given in Appendix~\ref{VLij}.  The Yukawa coupling matrix $y_{\ell}$ (\ref{g221e57}) is also mentioned,  with a requirement that  the SM block of the charged leptons is diagonal after the block-diagonalization. It does not affect results obtained in Ref.  \cite{g221}, which depend mainly on  the gauge couplings.

Hereafter, many calculations to discuss on phenomenology of  Higgs bosons will ignore  small mixing between different flavor quarks.  We will apply the same results of the charged lepton sector to the quarks.  The equivalences  between notations are: $V^{ij}_L, \lambda_{\ell}, \widetilde{M}_{L_{1,2}}, \Delta_{\mu,\tau}\rightarrow V^{ij}_{Q},\lambda_{q}, \widetilde{M}_{Q_{1,2}}, \Delta_{b,s}$, which were given in \cite{g221}.

 Next, we will discuss another possibility that neutrinos can get Majorana mass terms.
\subsection{Neutral lepton masses}
\label{neutrallepton}
   Unlike charged leptons, where the SM-like charged leptons have their own right-handed partners,
the SM-like neutrinos do not. In addition, the neutral leptons may inherit Majorana mass terms,
for example $\frac{1}{2}\overline{(\nu_L)^c}m_{\nu}\nu_L$ for active neutrinos.
Hence, it is more convenient to write the mass  matrix of neutral leptons in the form discussed
in the seesaw models~\cite{numixing}, which is different  from~\cite{g221}.
At the beginning $\nu_L$, $N_R$ and $N_L$ will be considered as independent fields, where the left-
and right-handed bases are $\mathcal{N}'^I_L=(\nu_L,\; (N_R)^c,\; N_L)^T$ and
$(\mathcal{N}'^{I}_L)^c=((\nu_L)^c,\; N_R,\; (N_L)^c)^T$, respectively.
The mass term in the Lagrangian is now
$-\mathcal{L}^{N}_{\mathrm{mass}}= \frac{1}{2} \overline{\mathcal{N}'_L} \mathcal{M}_N (\mathcal{N}'_L)^c
+ \mathrm{h.c.}$.
For $\mathrm{n_{VL}}=2$, the mass matrix of the neutral leptons is a $7\times 7$ symmetric matrix
having the following form:
\bea
\mathcal{M}^I_N=  \left(
                  \begin{array}{ccc}
                  0 & m_D &0 \\
                  m_D^T & 0 & M^T_L \\
                  0 & M_L & 0 \\
                  \end{array}
                  \right),
\label{NuMmatr}
\eea
where $m_D\equiv  \frac{1}{2} \lambda_\ell u$ and  $M_L$ are $3\times 2$ and $2\times 2$ matrices,
respectively. Similarly to seesaw models, \emph{$(N_L)^c$ and $N_R$ are additional right-handed neutrinos}.
The matrix~(\ref{NuMmatr}) can be generally diagonalized through the transformation
$\Omega^T \mathcal{M}^I_N \Omega=\mathrm{diag}(\widehat{m},\; \widehat{M})$,
where $\Omega$ is an unitary $7\times 7$ matrix;   $\widehat{m}=\mathrm{diag}(m_{\nu_1},m_{\nu_2},m_{\nu_3})$
are light Majorana neutrino masses and $\widehat{M}$ gives Dirac masses for heavy neutrinos.  Unfortunately, all light neutrinos are massless.
This can be proved as follows. The  neutral neutrino masses   are $m_I=\sqrt{x}$ where  the values of $x$ are  roots of the  equation, $\mathrm{det}\left[x\times I_7- \mathcal{M}^{I\dagger}_N \mathcal{M}^{I}_N\right]=0$.
For arbitrary forms of the $m_D$ and $M_L$, there are always three massless values of $x$.
The  matrices $M_L=\mathrm{diag}(M_{L_1},\;M_{L_2})$ and $\lambda_\ell$ in (\ref{laell1}) result in  four  other solutions: $x_4=x_5=u^2 \widetilde{M}^2_{L_1}$ and
$x_6=x_7=u^2 \widetilde{M}^2_{L_2}$. A pair of two degenerate values corresponds to one Dirac mass of
a heavy Dirac neutrino, the same as that mentioned in~\cite{g221}. The mixing matrix of neutrinos
is derived from (\ref{VL}) as follows:
\bea \Omega^T\equiv\left(
              \begin{array}{ccc}
                V^{11}_L& 0 & V^{12}_L \\
                0 & 1 & 0 \\
                V^{21}_L & 0 & V^{22}_L \\
              \end{array}
            \right) \rightarrow \Omega^T \mathcal{M}^I_N \Omega=\left(
                                                                  \begin{array}{ccc}
                                                                    0 & 0 & 0 \\
                                                                    0 & 0 & u \widetilde{M}_{L} \\
                                                                    0 & u \widetilde{M}_{L}& 0 \\
                                                                  \end{array}
                                                                \right),
\label{omega}
\eea
where new neutrino masses are pure Dirac. In addition, new lepton masses in each family
are nearly degenerate. Equation~(\ref{omega}) gives the relations between the original and mass
bases $ \mathcal{N}_{L}= V^{\dagger}_{L}\mathcal{N}^{(d)}_L$ and $\mathcal{N}_{R}=\mathcal{N}^{(d)}_R$,
which are the same as those of the charged leptons.

To keep the lepton spectrum being unchanged and looking for a solution of active neutrino mass problem,
the mass terms of active neutrinos must come from the effective  Majorana terms $\frac{1}{2}\overline{(\nu_L)^c}\; m_{\nu}\nu_L + \mathrm{h.c}$. Because the active neutrino masses are tiny,
their effect on the mixing parameters with heavy neutrinos is negligible.
Based on the mechanism of the neutrino mass generation in the Zee model~\cite{Zee},
in this model only one pair of new singly charged Higgs bosons, denoted as $\delta^{\pm}\sim (1,1,1)_{\pm1}$
carrying even $\mathcal{Z}_2$ charges, is introduced.
New couplings for generating one-loop radiative neutrino masses are
\bea
-\Delta \mathcal{L}&=& f_{ij}\overline{(\ell_{L_i})^c}(i\sigma_2) \ell_{L_j}\delta^+
+\sqrt{2}\lambda_{\delta}(i\sigma_2\phi')^T\Phi\phi\delta^- \crn
&+&  f^{\prime}_{kl}\overline{(L_{L_k})^c}(i\sigma_2) L_{L_l}\delta^+ +\mathrm{H.c.},
\label{Lnumass}\eea
where $i,j=1,2,3$; and  $k,l=1,2$. In  the general case, $k,l=1,2,..,n_{V_L}$. We stress that all terms in (\ref{Lnumass}) are simultaneously survival  only when both $\phi'$ and $\delta^{\pm}$ are  even under $\mathcal{Z}_2$ symmetry.

\begin{figure}[ht]
  \centering
  \includegraphics[width= 9cm]{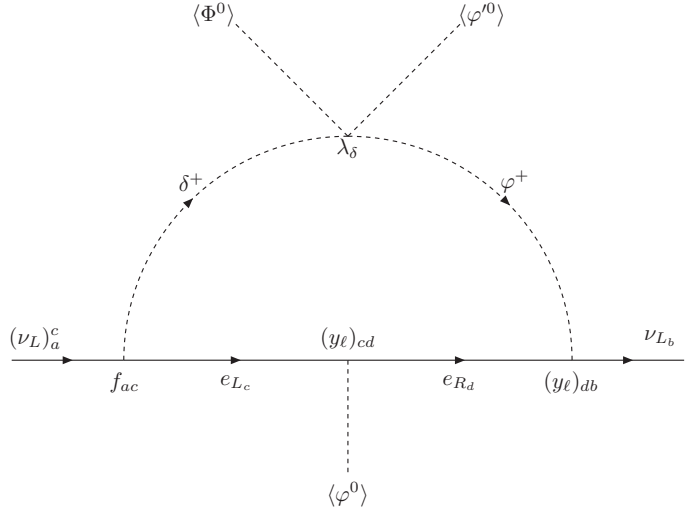}\\
  \caption{One loop correction to active neutrino masses}\label{acNuMass}
\end{figure}

The   terms in the first line of (\ref{Lnumass}) violate the lepton numbers, exactly in  the same way as in the Zee model,
where $\phi'$ plays a similar role as the second Higgs doublet. Similarly to the Zee model,
the trilinear coupling is $\lambda_{\delta}u$ after the first step of the spontaneous breaking.
An one-loop diagram generating active neutrino masses is shown in Fig.~\ref{acNuMass}.
Following~\cite{Zee,Znum}, the effective mass matrix of light neutrinos is derived in Appendix~\ref{zeem}, where $\varphi^\pm$ and $\delta^\pm$ are assumed to be
the physical Higgs bosons. But in the model under consideration, $\varphi^{\pm}$ and $\delta^\pm$
are not mass eigenstates. As we will discuss later, the physical  fields
 in the Higgs sector are $h^\pm_{1,2}$,
and there are some useful relations: $\varphi^\pm\sim c_{\zeta}h^\pm_1 $,
$\Phi^\pm \sim s_{\zeta}c_{\xi}h^\pm_1 -s_{\xi} h^\pm_2$, and $\delta^\pm\sim s_{\zeta}s_{\xi}h^\pm_1+c_{\xi} h^\pm_2$. The  parameters $c_{\xi},s_{\xi},c_{\zeta}$, and $s_{\zeta}$ involve with mixing parameters $\xi$ and $\zeta$ of the Higgs bosons,  defined in Eqs. (\ref{zangle}) and (\ref{xia}), as we will present
below.
The Higgs coupling in~(\ref{Lnumass}) can be rewritten as follows:
\bea
&& -\frac{1}{2}\lambda_{\delta} vc_{\beta}\left[ 2s_{\zeta}s_{\xi}\left(uc_{\zeta}
+vs_{\beta} s_{\zeta} s_{\xi}\right) h^{+}_1h^-_1  -2s_{\xi}c_{\xi}v  h^{+}_2h^-_2 \right. \crn
&+& \left. \left(uc_{\zeta}c_{\xi} +v s_{\beta} s_{\zeta}(c^2_{\xi}-s^2_{\xi}) \right)
\left(h^{+}_2h^-_1+\mathrm{H.c.} \right) \right]  \subset
-\sqrt{2}\lambda_{\delta}(i\sigma_2\phi')^T\Phi\phi\delta^- +\mathrm{H.c.},
\label{laxcoup}
\eea
where $v_{\phi}=v s_{\beta}$,  $v_{\phi'}=vc_{\beta}$ and $t_{\beta}\equiv\tan\beta=s_{\beta}/c_{\beta}$, which are defined in~\cite{g221}.

 Charged leptons $e_c$ in the loop will be considered as mass eigenstates with masses $m_{e_c}$.
Therefore, the Yukawa terms should be written in terms of physical charged lepton states,
and light neutrinos are massless states after the rotation $V_L$.  For simplicity, we will assume that
$V^{11}_L$ is real and  $V_e,W_e\simeq I$. In addition, we ignore one-loop contributions to heavy neutrino
masses because they are extremely smaller than the tree level masses. Then the one-loop corrections  are  mainly from light leptons, namely
\be
f_{ac}\overline{e_{L_c}}(\nu_{L_a})^c=f_{ac}(V_L)_{Ih}(V^T_L)_{cJ}\overline{\mathcal{E}^{(\mathrm{d})I}_{L}}
(\mathcal{N}^J_{L})^c\rightarrow f_{gh}(V_L)_{ch}(V^T_L)_{ga}\overline{e_{L_c}}(\nu_{L_a})^c
=(V^{11}_LfV^{11}_L)_{ac}\overline{e_{L_c}}(\nu_{L_a})^c,
\nn
\ee
and $
f^{\prime}_{kl}\overline{L_{L_k}}(N_{L_l})^c\rightarrow (V^{12}_Lf^{\prime}V^{12T}_L)_{ac}\overline{e_{L_c}}(\nu_{L_a})^c$. Here  $a,c,g,h=1,2,3$; $I,J=1,2,...,7$;  $f$  and $f'$ are  $3\times3$ and $2\times2$ antisymmetric matrices, respectively. Similarly, we have
\bea
(y_{\ell})_{bc}\overline{\nu_{L_b}}e_{R_c}&=&\frac{\sqrt{2}}{vs_{\beta}}\left(\mathcal{M}_{\mathcal{E}}\right)_{bc}
(V_L)_{Jb} \overline{\mathcal{N}^J_{L}}e_{R_c}\simeq \frac{\sqrt{2}}{vs_{\beta}}\left(V^{\dagger}_L
\mathrm{diag}(m_{\mathcal{E}_I})\right)_{bc} (V_L)_{Jb} \overline{\mathcal{N}^J_{L}}e_{R_c},
\crn
&\rightarrow&  \frac{\sqrt{2}}{vs_{\beta}}(V^{T}_L)_{dc}(V_L)_{bd}m_{e_c} \overline{\nu_{L_b}}e_{R_c}
= \frac{\sqrt{2}}{vs_{\beta}}(V_LV^{T}_L)_{bc}m_{e_c} \overline{\nu_{L_b}}e_{R_c}.
\nn
\eea
The effective mass matrix $m_{\nu}$ of active neutrinos is derived  based on~(\ref{mnu2}),
\bea
(m_{\nu})_{ba} &=&\frac{\lambda_{\delta}\sqrt{2}}{16\pi^2t_{\beta}}\times  \frac{u }{m^2_{h^\pm_1}}
\sum_{c=1}^{3}m^2_{e_c}\left[(V^{11}_LfV^{11}_L+  V^{12}_Lf^{\prime}V^{12T}_L)_{ac}(V^{11}_LV^{11T}_L)_{bc} \right.\crn
 && \left. +(V^{11}_LfV^{11}_L+ V^{12}_Lf^{\prime}V^{12T}_L)_{bc}
(V^{11}_LV^{11T}_L)_{ac}\right]
\crn
 &\times&\left(\left[c_{\zeta} c_{\xi}+\frac{v}{u}s_{\beta}s_{\zeta}(c^2_{\xi}-s^2_{\xi})\right]
\frac{m^2_{h^\pm_1}}{m^2_{h^\pm_1}-m^2_{h^\pm_2}} \ln\left[\frac{m^2_{h^\pm_2}}{m^2_{h^\pm_1}}\right]
- 2s_{\zeta}s_{\xi}\left[ c_{\zeta}+\frac{v}{u}s_{\beta}s_{\zeta}s_{\xi}\right]\right)\crn
 &\equiv& \frac{1}{m_0}\left\{\left(\left[V^{11}_LfV^{11}_L+V^{12}_Lf^{\prime}V^{12T}_L\right] M^2_eV^{11T}_LV^{11}_L\right)_{ab}
\right. \crn
&+& \left.\left(\left[V^{11}_LfV^{11}_L+V^{12}_Lf^{\prime}V^{12T}_L\right] M^2_eV^{11T}_LV^{11}_L\right)_{ba}\right\}, \label{mnug221}
\eea
where $M_e\equiv \mathrm{diag}(m_{e},\;m_{\mu},\;m_{\tau})$.

Including loop contributions (\ref{mnug221}), the SM block of (\ref{omega}) will be  changed from zero into $m_{\nu}$: $\Omega^T \mathcal{M}^I_N \Omega\supset m_{\nu}=U^*_{\mathrm{PMNS}}\hat{m}_{\nu} U^\dag_{\mathrm{PMNS}}$, where $\hat{m}_{\nu}=\mathrm{diag}(m_{\nu_1},\,m_{\nu_3},\,m_{\nu_3})$ consisting of three active neutrino masses, and $U_{\mathrm{PMNS}}$ is the well-known neutrino mixing matrix.  If $V^{11}_L=I_3$ and $V^{12}_L=0$, the Eq. (\ref{mnug221}) is the same (as~ \ref{mnu2}).  Like in the Zee models,
the parameters arising from the Higgs sector affect the order of the neutrino masses only. But the  masses and mixing angles  of active neutrinos depend on unknown parameters in $f, f^{\prime}$, and $V^{11}_L$.   As a result, the  model under consideration is less restrictive  in fitting the neutrino data  than the Zee models. Because these models are still valid~\cite{zeenudata}, the neutrino sector mentioned here is realistic.  In general, fitting recent neutrino data needs at least five free parameters, in agreement with three mixing angles and two squared mass differences.
 Because  two of four parameters, namely  $m_0$ and three $f_{ab}$, determine the order of the lightest neutrino mass,
there are two free parameters left.   When  $n_{V_L}\geq 3$, there are at least three additional parameters $f^{\prime}_{kl}$, enough for fitting neutrino data without constraints on $V^{11}_L$. Interestingly, the neutrino fitting results  in \cite{zeenudata} would be applied to the model under consider ration if $L_L$ carries even $\mathcal{Z}_2$ charge which will survive  the  lepton coupling matrix $\tilde{y}_{\ell}$ in (\ref{g221e59}). 

Regarding  $n_{V_L}=2$, there is only one   parameter $f'_{12}=-f'_{21}$.  Therefore two parameters in  $V^{11}_L$ may be involved with fitting neutrino data.  Our numerical investigation showed that the allowed regions in Ref. \cite{g221},  controlled by  the texture $\lambda_{\ell}$  (\ref{laell1}),  seems much more constrained. Note that the $m_{\nu}$ in (\ref{mnug221}) keep only main contributions from  loops containing light charged leptons, where mixing terms with order $\mathcal{O}(\epsilon)$ are ignored. With $u$ around 1 TeV and  light new charged leptons, contributions from these lepton mediations to $m_{\nu}$ will be significant, implying that their masses can be free parameters for fitting neutrinos data without much changes of  $\Delta_{\mu,\tau}$. Finding exact  allowed  regions should be done elsewhere.

When the neutrino data is fitted, the results in Ref. \cite{g221} for B-decay  anomalies are still unchanged because the analysis considered here addresses only effects of  tree contributions from heavy gauge bosons, where other contributions from the light lepton masses are suppressed.  The unique  changes may come from the gauge couplings of active neutrinos with charged gauge bosons. Following  \cite{g221}, after the block-diagonalization these gauge couplings are proportional to  $W^{\mu}_l\overline{ \nu_L}\gamma_{\mu}e_L$ and $W^{\mu}_h\overline{ \nu_L}\gamma_{\mu}\Delta^{\ell}e_L$, where $W_l$ and $W_h$ are light and heavy charged gauge bosons. In the neutrino mass basis  they become  $W^{\mu}_l\overline{ \nu_L}U^{\dagger}_{\mathrm{PMNS}}\gamma_{\mu}e_L$ and $W^{\mu}_h\overline{ \nu_L}U^{\dagger}_{\mathrm{PMNS}}\Delta^{\ell}\gamma_{\mu}e_L$,  resulting in the same factor  $(U^{\dagger}_{\mathrm{PMNS}})_{ii}$ for coupling $ \overline{\nu_i}e_{i}$ with a diagonal $\Delta^{\ell}$ obtained from the texture of $\lambda_{\ell}$ in (\ref{laell1}).  This factor will not appear in final results presenting the ratios  of  B-decay anomalies, as given in \cite{g221}.

In general, active neutrino mass generation from radiative corrections mentioned above affects only the lepton sector. Furthermore, it does not affect mixing parameters  controlling the $\lambda_{\ell}$ structure at the first breaking step, hence suggests that the orders of numerical values in allowed regions will not change after neutrino data is fitted.

The above discussion just refers to a simple extension that can generate active neutrino masses
through radiative corrections. The problems of neutrino masses and DM can be solved by models
with more charged Higgs bosons and singlet right-handed neutral leptons, such as~\cite{nuDM}.
Following the structures of these models, apart from $\delta^\pm$, at least one pair of singly
charged Higgs bosons $S^{\pm}$ and a neutral lepton $\mathcal{F}_R\sim (1,1,1)_{0}$ have to be introduced,
where $S^+\sim (1,1,1)_{1}$. In addition, only $S^{\pm}$ and $\mathcal{\mathcal{F}}_R$ are odd under
a new $Z_2$ discrete symmetry,
$\left\{ S^\pm,\; \mathcal{F}_R\right\}\rightarrow \left\{ -S^\pm,\; -\mathcal{F}_R\right\}$.
Therefore,  $\mathcal{F}_R $ can play the role of DM. It has a  Majorana mass term of the form
$ \frac{1}{2} \overline{(\mathcal{F}_R)^c}m_{\mathcal{F}}\mathcal{F}_R$.
Active neutrinos get mass from loop corrections, which arise from a new Yukawa term,
$-\Delta\mathcal{L}_Y=f_{ij}\overline{(\ell_{L_i})^c }(i\sigma_2)\ell_{L_j}\delta^+
+g_{i}\overline{(\mathcal{F}_R)^c }e_{R_i}\delta^+ +\mathrm{H.c.}$,
and a coupling of charged Higgs bosons,
$ \frac{1}{4}\lambda_{\delta S} (\delta^+)^2 (S^-)^2+\mathrm{ H.c.}$.
This kind of models seems to be less interesting because the origin of neutrino masses is not related
to the new leptons.

New ingredients for generating radiative corrections to active neutrinos do not change both results
of the gauge sector and LNU discussed in~\cite{g221}, because no new breaking scale contributes to
the masses of gauge Higgs bosons. If we add a new $SU(2)_1$ triplet, denoted as $\Delta\sim (1,3,1)_{1}$,
creating an Yukawa term like $ -Y_{\Delta}\overline{(L_L)^c}i\sigma_2\Delta L_L+ \mathrm{H.c.}$,
a neutral component of this triplet will develop a non-zero VEV $v_{\Delta}$, which contributes a new mass
term of the form $ \frac{1}{2}\mu_X \overline{N_L} (N_L)^c+ \mathrm{H.c.}$
to the neutrino mass matrix~(\ref{NuMmatr}). This matrix has the same form shown in the inverse seesaw
models~\cite{numixing,iss}. Hence the active neutrino masses will be non-zero. In addition, some new
neutrinos may get light masses and play the role of DM~\cite{DMiss}. These models seem interesting
because they may give connections between the $SU(2)_1$ leptons with neutrino masses and DM.
But the appearance of the new vev $v_{\Delta}$ will contribute to masses and mixing parameters of
the Higgs and gauge bosons, consequently it will affect the results shown in~\cite{g221}.
This extension is beyond our scope, and  should be thoroughly studied in another work.

Now we turn to one of the most important elements: gauge bosons.
\subsection{Gauge boson masses}
\label{gaugeboson}
Gauge boson masses arise from the piece
\be
\mathcal{L}_{\mathrm{gauge\; boson\; mass}} = (D_\mu \langle \phi \rangle)^\dag D^\mu \langle \phi \rangle
+ (D_\mu \langle \phi' \rangle)^\dag D^\mu \langle \phi' \rangle
+ {\rm Tr}[(D_\mu \langle \Phi \rangle)^\dag D^\mu \langle \Phi \rangle]\, ,
\label{g221e12}
\ee
where the covariant derivative of $\Phi$ is determined as
\bea
\left(D_\mu \Phi\right)_\al ^\bet & = &  \pa_\mu \Phi_\al^\bet
- \fr i 2  g_1 W_{i \mu}^1 (\si_i)_\al^\ga (\Phi)_\ga^\bet
+ \fr i 2  g_2 (\Phi)_\al^\ga  W_{i \mu}^2 (\si_i)_\ga^\bet\, .
\label{g221e13}
\eea
With the help of the notation
\be
W_\mu^i \equiv \fr 1 2 \sum_{\al=1}^3 W_{\al \mu}^i \si_\al = \fr 1 2 \left(%
\begin{array}{cc}
W_3^i & \sqrt{2}\,  W_{i}^+ \\
\sqrt{2}\,  W_{i}^-
& -W_3^i \\
\end{array}\,
\right)_\mu \, ,  W_i^{ \pm} \equiv  \fr{1}{\sqrt{2}}(W_1^i \mp i W_2^i)\, , i = 1,2\, ,
\label{g221e14}
\ee
 contributions to masses of gauge bosons are
\bea
{\rm Tr}[(D_\mu \langle \Phi \rangle)^\dag D^\mu \langle \Phi \rangle]
&=& \fr{u^2}{16}\left[ 2 (g_1 W_3^1 - g_2 W_3^2)^2 + 4
(g_1 W_{1}^+ - g_2 W_{2}^+)_\mu ( g_1 W_{1}^- - g_2 W_{2}^-)^\mu\right]\, ,\crn
(D_\mu \langle \phi \rangle)^\dag D^\mu \langle \phi \rangle
& = & \fr{v_\phi^2}{4} \left[ g_2^2 W_2^{ +}W_2^{ -} + \fr 1 2 (g_2 W_2^3 - g' B)^2\right]\, ,\crn
(D_\mu \langle \phi' \rangle)^\dag D^\mu \langle \phi' \rangle
& = & \fr{v_{\phi'}^2}{4} \left[ g_1^2 W_1^{ +}W_1^{ -} + \fr 1 2 (g_1 W_1^3 - g' B)^2\right]\, .\nn
\label{g221e17}
\eea
From this, masses and eigenstates of gauge bosons can be found in agreement  with those presented in Ref. \cite{g221}.  We will review important aspect then  discuss some new properties when masses and mixing angles are calculated up to order of $\mathcal{O}(\epsilon^2)$.
\subsection{Neutral gauge bosons}
In the basis $(W_3^1, W_3^2, B)$ the squared mass matrix of neutral gauge boson is  $ M^2_{nb}$. At the first step, where $v_{\phi},v_{\phi'}\rightarrow0$,  only two states $W_3^1$ and $ W_3^2$ are rotated through a rotation $ C_1$ so that $ C_1  M^2_{nb} C_1^T|_{v_{\phi},v_{\phi'}=0}
=\mathrm{Diag}\left(0,0,\frac{1}{4}(g_1^2+g_2^2)u^2\right)$. Elements of $C_1$ depend on a mixing angle $\beta'$ defined by
\be
\tan\beta'\equiv \frac{g_1}{g_2}, \; c_{\beta'}=\cos\beta' =g_2/n_1,\;s_{\beta'}=\sin\beta'=g_1/n_1,
\label{betap}
\ee
where  $n_1=\sqrt{n_1^2+n_2^2}$ was used already in~\cite{g221}.

The first breaking step implies the following transformation of the neutral gauge bosons: $(W^1_3, W^2_3, B)\xrightarrow{u}( B, W_3, Z_h)$,  where $(B, W_3)$ are the SM gauge bosons.
We have  $(W^1_3, W^2_3, B)^T=C_1^T (B, W_3,Z_h )^T$, i.e.  $W_3= c_{\beta'} W^1_3 +s_{\beta'} W^2_3 $ and $Z_h= s_{\beta'} W^1_3 -c_{\beta'} W^2_3$, where  $g=g_1g_2/n_1$ and $g'$ are identified as the SM gauge couplings;
$v_{\phi}= vs_{\beta}$,  $v_{\phi'}=vc_{\beta}$ \cite{g221}. Note that $v\simeq 246$ GeV,  $g'=gs_W/c_W$, and
$s_W$ is the sine of the Weinberg angle. From now on, $n_1, g_1,g_2$ and $n_2$ will be written as
\be
n_1= \frac{g}{c_{\beta'}s_{\beta'}}, \hs n_2=\frac{g}{c_W}, \hs
g_1=  \frac{g}{c_{\beta'}}, \hs g_2=  \frac{g}{s_{\beta'}}.
\label{newnotation}
\ee
At the second step, the mixing matrix $C_2$ is the SM rotation of only $B$ and $W^3$, giving new basis $(A,Z_l,Z_h)^T=C_2 (Z_h, W_3, B)^T= C_2C_1(W^1_3, W^2_3, B)^T $, where $A$ and $Z_l$ are the photon and SM gauge boson.  The  respective  matrix $M'^2_{nb}$  is
\bea
M'^2_{nb}=C_2 C_1 M^2_{nb}(C_2C_1)^T= \frac{g^2}{4} \left(%
\begin{array}{ccc}
	0  & 0  & 0 \\
	0 &\frac{v^2}{c^2_W} &\frac{\left(c_{2\beta}-c_{2\beta'}\right)v^2}{c_Ws_{2\beta'}}  \\
	0 &\frac{\left(c_{2\beta}-c_{2\beta'}\right)v^2}{c_Ws_{2\beta'}}
	& \frac{4 u^2+\left(1-2c_{2\beta}c_{2\beta'}+ c^2_{2\beta'}\right) v^2}{s^2_{2\beta'}} \\
\end{array}\,
\right).
\label{g221e22}
\eea
The mass eigenstates $(Z,Z')$ relates with the $Z_l-Z_h$ mixing angle defined  as
\bea
t_{2Z}\equiv\tan(2Z) &=& \frac{-2 \left( M'^2_{nb}\right)_{23}}{\left( M'^2_{nb}\right)_{33}
-\left( M'^2_{nb}\right)_{22}}
= \frac{2 s_{2\beta'}\left(c_{2\beta'}-c_{2\beta}\right)\frac{\epsilon^2}{c_W}}{4+\left(1-2c_{2\beta}c_{2\beta'}
+c^2_{2\beta'}-\frac{s^2_{2\beta'}}{c^2_W}\right)\epsilon^2},
\label{xiZZp}
\eea
where $\epsilon \equiv \frac{v}{u}$. The $Z_l-Z_h$ mixing vanishes when $\beta'=\beta$, where $\tan\beta=s_{\beta}/c_{\beta}$.

The masses of the physical eigenstates $(Z,Z')$ are
\bea
M_Z^2& =&\fr{g^2}{4}\left[\frac{v^2}{c^2_W}c^2_{Z}+\left( \frac{4 u^2+\left(1-2c_{2\beta}c_{2\beta'}
+ c^2_{2\beta'}\right) v^2}{s^2_{2\beta'}}\right) s^2_{Z}
+ 2\left( \frac{\left(c_{2\beta}-c_{2\beta'}\right)v^2}{c_W s_{2\beta'}} \right)s_{2Z} \right],
\crn
M_{Z'}^2 &=&\fr{g^2}{4}\left[\frac{v^2}{c^2_W}s^2_{Z}
+\left( \frac{4 u^2+\left(1-2c_{2\beta}c_{2\beta'}+ c^2_{2\beta'}\right) v^2}{s^2_{2\beta'}}\right) c^2_{Z}
- 2\left( \frac{\left(c_{2\beta}-c_{2\beta'}\right)v^2}{c_W s_{2\beta'}} \right)s_{2Z} \right]  . \nn
\label{mzz'}
\eea
The relation between the two bases $(W_1,W_2, B)$ and $(A,Z,Z')$ is
\bea
\left(%
\begin{array}{c}
W_1\\
W_2\\
B\\
\end{array}\,
\right) &=&
\left(%
\begin{array}{ccc}
s_Wc_{\beta'} &c_Wc_{\beta'}  &s_{\beta'} \\
s_Ws_{\beta'}&c_Ws _{\beta'} &-c_{\beta'} \\
c_W &- s_W & 0\\
\end{array}\,
\right)
\left(
  \begin{array}{ccc}
    1 & 0 & 0 \\
    0 & c_Z & -s_Z \\
    0 & s_Z & c_Z \\
  \end{array}
\right)
\left(%
\begin{array}{c}
A\\
Z\\
Z'\\
\end{array}\,
\right),
\label{g221e29}
\eea
where  $(C_2C_1)^T$ is the first matrix in the right hand side of (\ref{g221e29}).

Using the new notations of~(\ref{newnotation}), the parameter $\zeta$ in~\cite{g221} can be expressed as
$\zeta\equiv s^2_{\beta}-\frac{g_1^2}{g_2^2}c^2_{\beta}= \frac{1}{2c^2_{\beta}} \left( c_{2\beta'}- c_{2\beta}\right)$.
In addition, from
$ m^2_{Z}\simeq \frac{g^2}{4 c^2_W}v^2\simeq \frac{g^2}{4 c^2_W} u^2 \epsilon^2$, $ m^2_{Z'}\simeq \frac{g^2}{4 c^2_{\beta'}s^2_{\beta'}}u^2= \frac{g^2}{ s^2_{2\beta'}}u^2$
and $\frac{g}{n_2}\frac{g_2}{g_1}=\frac{c_{\beta'}c_W}{s_{\beta'}}$, we can deduce an approximate form  $\xi_Z\simeq 1/2\tan 2\xi_Z$ in the limit $\epsilon\ll 1$, consistent with the expression of $\tan 2\xi_Z$
shown in (\ref{xiZZp}).
\subsection{Charged gauge bosons}
In the basis $(W_1^+, W_2^+)$ the squared  mass matrix of  charged gauge bosons was given in Ref \cite{g221}. Setting $v=0$, we can define a new basis: $W^+_l = ( c_{\beta'} W^+_1 +  s_{\beta'} W_2^+)$ and $W^+_h = (s_{\beta'} W^+_1 -  c_{\beta'} W_2^+)$, where the corresponding  squared mass matrix  is
\be
M_c^2 =  \frac{g^2}{4} \left(%
\begin{array}{cc}
v^2  & \frac{\left(c_{2\beta}-c_{2\beta'}\right)v^2}{s_{2\beta'}} \\
\frac{\left(c_{2\beta}-c_{2\beta'}\right)v^2}{s_{2\beta'}} & \frac{4 u^2+\left(1-2c_{2\beta}c_{2\beta'}+ c^2_{2\beta'}\right) v^2}{s^2_{2\beta'}} \\
\end{array}\,
\right).
\label{g221e32}
\ee
The SM-like boson $W^\pm$ is identified with $W^\pm\equiv W^\pm_l$ with mass $m_{W_l}= gv/2$.
The mixing $W^+_l-W^+_h$ is defined through the mixing angle $\xi_W$ satisfying
\bea
t_{2\xi_W}\equiv\tan(2\xi_W) &=&\frac{-2 \left( M^2_{c}\right)_{12}}{\left(M^2_{c}\right)_{22}
-\left(M^2_{c}\right)_{11}}
=  \frac{2 s_{2\beta'}\left(c_{2\beta'}-c_{2\beta}\right)\epsilon^2}{4+\left(1-2c_{2\beta}c_{2\beta'}
+ c_{4\beta'}\right)\epsilon^2} = c_W t_{2Z}\, .
\label{xiWa}
\eea
From (\ref{xiWa}), it follows that the ratio of the tangents of $W-W'$ and $Z-Z'$ mixing angles is $c_W$.
This will \emph{reduce the number of parameters in the model by 1}.

The physical mass eigenstates $(W^\pm, W'^\pm)$ are given by
\be \left(%
\begin{array}{c}
W^\pm\\
W'^\pm\
\end{array}\,
\right) =  \left(%
\begin{array}{cc}
c_{\xi_W}  & s_{\xi_W} \\
-s_{\xi_W}& c_{\xi_W}\\
\end{array}\,
\right)\left(%
\begin{array}{c}
W^\pm_l\\
W^\pm_h\
\end{array}\,
\right)
\label{g221e34}
\ee
with $c_{\xi_{W}}\equiv\cos\xi_W$, $s_{\xi_{W}}\equiv\sin\xi_W$,  and  masses
\bea
M_W^2& =&\fr{g^2}{4}\left[v^2c^2_{\xi_W}+\left( \frac{4 u^2+\left(1-2c_{2\beta}c_{2\beta'}
+ c^2_{2\beta'}\right) v^2}{s^2_{2\beta'}}\right)s^2_{\xi_W}
+ 2\left( \frac{\left(c_{2\beta}-c_{2\beta'}\right)v^2}{s_{2\beta'}} \right)s_{2\xi_W} \right],
\crn
M_{W'}^2 &=&  \fr{g^2}{4}\left[v^2s^2_{\xi_W}+\left( \frac{4 u^2+\left(1-2c_{2\beta}c_{2\beta'}
+ c^2_{2\beta'}\right) v^2}{s^2_{2\beta'}}\right)
c^2_{\xi_W}- 2\left( \frac{\left(c_{2\beta}-c_{2\beta'}\right)v^2}{s_{2\beta'}} \right)s_{2\xi_W} \right] .\crn
\label{g221e35p}
\eea
Note that $Z$ and $W$ are the SM-like gauge bosons.

We will derive the approximate formulas for the mixing angles and masses of the SM-like gauge boson
up to the order of $ v^2\times \mathcal{O}(\epsilon^2)$ because the corrections at this order
to the masses may contribute significantly to precision tests such as the $\rho$ parameter.  Because $t_{2Z}, t_{2\xi_W}\sim \epsilon^2$, we have
$s_{Z}\simeq t_{2Z}/2 $, $s_{\xi_W}\simeq t_{2\xi_W}/2 $. From (\ref{xiZZp}) and (\ref{xiWa}), we get
\bea
s_Z \simeq t_Z\simeq  \frac{s_{2\beta'}  \left(c_{2\beta}- c_{2\beta} \right)}{4 c_W} \epsilon^2, \;
s_{\xi_W} \simeq s_Z c_W .
\label{appgmixing}
\eea
This means that $s_Z^2,s^2_{\xi_W}\sim \epsilon^4$, hence
$c_Z^2=1-s_Z^2=1- \mathcal{O}(\epsilon^4)=1, \; c^2_{\xi_W}=1$.
For this reason, the masses of the gauge bosons in (\ref{mzz'}) and (\ref{g221e35p}) can be written as
\bea
M^2_{Z}&\simeq& \frac{g^2v^2}{4 c^2_W} \left[1+ \frac{(c_{2\beta}-c_{2\beta'})^2 \epsilon^2}{4} \right],
\hs M^2_{W}\simeq \frac{g^2v^2}{4} \left[1+ \frac{(c_{2\beta}-c_{2\beta'})^2 \epsilon^2}{4} \right],
\crn
M_{Z'}^2 &\simeq&M_{W'}^2 \simeq  \fr{g^2}{4} \frac{4 u^2+\left(1-2c_{2\beta}c_{2\beta'}
+ c^2_{2\beta'}\right) v^2}{s^2_{2\beta'}}.
\label{apgmass}
\eea
Then we have $\frac{M^2_W}{M^2_{W'}}\simeq \frac{s^2_{2\beta'}\epsilon^2}{4 }\simeq c_W^2\frac{M_Z^2}{M^2_{Z'}}$. In addition, at the tree level the $\rho$ parameter satisfies
$\rho=\frac{M^2_W}{c^2_W M^2_Z}=1+\mathcal{O}\left(\frac{v^4}{u^4}\right)$.
Hence, a TeV scale of $u$ gives a contribution to
$(\rho-1)= \mathcal{O}\left(\frac{v^4}{u^4}\right)\sim10^{-4}$,
in agreement with the recent experimental constraint on $\rho$ parameter~\cite{pdg2016}.
\section{Currents}
\label{current}
The Lagrangian $L_{\mathrm{fermion}} =  i \sum_f {\bar f} \ga^\mu D_\mu f$ contains interactions of the gauge bosons with the fermions. Let us firstly consider neutral currents. From  Eq.~(\ref{g221e29}), one gets
\bea
L_{NC}
& = &   \sum_f {\bar f} \ga_\mu A^\mu \left\{ g' c_W Y + \fr{g_1 g_2 s_W}{n_1}(T_3^1 + T_3^2)\right\} f
 \label{g221e37}\\
&+&  \sum_f {\bar f} \ga_\mu Z_l^\mu \left\{ -g' s_W Y + \fr{g_1 g_2 c_W}{n_1}(T_3^1 + T_3^2)\right\} f
\label{g221e38}\\
&+&  \sum_f {\bar f} \ga_\mu Z_h^\mu \left(\fr{g_1^2  }{n_1}T_3^1 - \fr{g_2^2  }{n_1} T_3^2\right) f.
\label{g221e39}
\eea
Using  $e = g' c_W = g s_W$,  etc.,  expression in~(\ref{g221e37})
gives the well-known  electromagnetic current
$ L_{em} =  A_\mu J_{em}^\mu
= e A_\mu  \sum_f {\bar f} \ga^\mu   Q f $,
where $Q$ is the electric charge operator defined in~(\ref{g221e7}).

Neutral currents are  defined as $L_{NC}^{Z_l,Z_h} =  Z_l^\mu J_\mu (Z_l)+ Z_h^\mu J^\mu(Z_h)$, where  $J_\mu (Z_l)$ and $J^\mu(Z_h)$ can be found  from
(\ref{g221e38}) and (\ref{g221e39}).
 Remind that physical neutral gauge bosons are $Z$ and $Z'$ defined from the $Z_l-Z_h$
mixing angle~(\ref{xiZZp}), leading to the respective  neutral currents
\be
J^\mu(Z)  = \fr{ c_Z g}{c_W} \sum_f {\bar f} \ga^\mu  \left(T_3^1 + T_3^2 - s^2_W  Q \right)  f
+ \frac{ s_Z g}{t_{\beta'}} \sum_f {\bar f} \ga^\mu \left(t^2_{\bet '} T_3^1 -  T_3^2)\right) f\, ,
 \label{g221e46}
\ee
and
\be
J^\mu(Z')  = \frac{ c_Z g}{t_{\beta'}}  \sum_f {\bar f} \ga^\mu \left(t^2_{\bet '}T_3^1 -  T_3^2)\right) f
-  \fr{ s_Z g}{c_W} \sum_f {\bar f} \ga^\mu  \left(T_3^1 + T_3^2 - s^2_W  Q \right)  f\, .
\label{g221e47}
\ee
The second term in (\ref{g221e46}) is the NP contribution.

Let us write explicitly the neutral current of the $Z$ boson
\bea
J^\mu(Z) & = & \fr{ c_Z g}{c_W} \sum_{\psi= q,\ell} {\bar \psi} \ga_\mu  \left( T_3^2 - s^2_W  Q \right)
\psi \hs \textrm{(as SM)}\crn
&-&
\frac{g s_Z  }{2t_{\beta'}}  \left({\bar \nu}_L \ga_\mu    \nu_L  + {\bar u}_L \ga_\mu    u_L
 - {\bar l}_L \ga_\mu    l_L  - {\bar d}_L \ga_\mu    d_L\right)\crn
& + & \fr{ c_Z g}{ c_W} \left[  \fr 1 2   {\bar N} \ga_\mu  N +  {\bar E} \ga_\mu
\left( -\fr 1 2 +  s^2_W   \right)  E\right.\crn
&+&
\left. {\bar U} \ga_\mu  \left( \fr 1 2 - \fr 2 3 s^2_W   \right)  U +  {\bar D} \ga_\mu
\left( -\fr 1 2 + \fr 1 3 s^2_W   \right)  D\right]\crn
&+&
 \frac{g s_Z  t_{\beta'}}{2}\left( {\bar N} \ga_\mu  N -   {\bar E} \ga_\mu  E  + {\bar U} \ga_\mu  U
-   {\bar D} \ga_\mu  D\right)\, ,
\label{g221e46t}
\eea
where $T^1_3=0$ and $T^2_3=\frac{1}{2}\sigma_3$ for the SM fermion  doublets,
and $T^1_3=\frac{1}{2}\sigma_3, T^2_3=0$  for the extra fermion doublets. Only interactions in the first line of  (\ref{g221e46t})  are the SM ones.  The remaining
provides NP effects. Note that interactions of new vector-like fermions include both $P_L$ and $P_R$
parts (as vector).

Let us write the couplings of the $Z$ boson with  physical fermion states  in the form
\be
L^{NC}(Z,f) = \fr{ c_Z g}{c_W} Z_\mu {\bar f}\ga^\mu (g_L  P_L + g_R P_R) f\, ,
\label{g221e46t1}
\ee
where $P_{L,R} = (1\mp \ga_5)/2$.
The couplings $g_L$ and $g_R$ are listed in Table~\ref{tZff}, where we denote
\be
\rho_{\mu\tau}\equiv\sqrt{1-c^2_{\beta'}(\Delta^2_{\mu}+\Delta^2_{\tau})},\hs
\rho_{sb}\equiv\sqrt{1-c^2_{\beta'}(\Delta^2_{s}+\Delta^2_{b})}.\label{deltalq}
\ee
\begin{table}[h]
 \centering
 \caption{ Couplings of $Z$ boson with fermions, $ \ell=e,\mu,\tau$ and $\Delta_e=\Delta_d=1$. }\label{tZff}
 \begin{tabular}{|c|c|c|c|c|}  \hline
f &$ g_L $ &$ g_R $ & $g_V $ & $ g_A$ \\
\hline
$\nu_{\ell}$ &$\fr 1 2 -\fr{\left(1-\Delta^2_{\ell}\right)c_W t_Z }{2t_{\bet'}}$ & $0$&$\fr 1 2 -\fr{\left(1-\Delta^2_{\ell}\right)c_W t_Z }{2t_{\bet'}}$ &$\fr 1 2 -\fr{\left(1-\Delta^2_{\ell}\right)c_W t_Z }{2t_{\bet'}}$ \\
\hline
$\ell=e, \mu, \tau $&$- \fr 1 2 + s_W^2 + \fr{\left(1-\Delta^2_{\ell}\right)c_W t_Z }{2t_{\bet'}} $  &$s_W^2$ & $-\fr 1 2  + 2s_W^2+ \fr{\left(1-\Delta^2_{\ell}\right)c_W t_Z }{2t_{\bet'}}$&$ -\fr 1 2 +\fr{\left(1-\Delta^2_{\ell}\right)c_W t_Z }{2t_{\bet'}}$ \\
\hline
$\overline{\mu}\tau,\;\overline{\tau}\mu$&$ -\fr{\Delta_{\mu}\Delta_{\tau}c_W t_Z }{2t_{\bet'}} $  &$0$ & $ -\fr{\Delta_{\mu}\Delta_{\tau}c_W t_Z }{2t_{\bet'}} $  &$ -\fr{\Delta_{\mu}\Delta_{\tau}c_W t_Z }{2t_{\bet'}}$  \\
\hline
$q=u, c, t$ &$\fr 1 2 -\fr 2 3 s_W^2 -\fr{(1-\Delta^2_{q})c_W t_Z }{2t_{\bet'}}$ &$-\fr 2 3 s_W^2 $& $\fr{1}{2} -\fr 4 3 s_W^2  -\fr{(1-\Delta^2_{q})c_W t_Z }{2t_{\bet'}} $&$\fr 1 2 -\fr{(1-\Delta^2_{q})c_W t_Z }{2t_{\bet'}}$\\
 \hline
$q=d, s, b$ &$-\fr 1 2 + \fr 1 3 s_W^2 + \fr{(1-\Delta^2_{q})c_W t_Z }{2t_{\bet'}}$ &$ \fr 1 3 s_W^2 $& $  -\fr 1 2 + \fr 2 3 s_W^2 +\fr{(1-\Delta^2_{q})c_W t_Z }{2t_{\bet'}}$&$-\fr 1 2 +\fr{(1-\Delta^2_{q})c_W t_Z }{2t_{\bet'}}$\\
 \hline
$N_1$&$\fr 1 2 - \fr{c_W t_Z }{t_{2\bet'}}$&$\fr 1 2 +\fr{c_W t_Z t_{\bet'}}{2} $&$1+ \fr{c_W(1-3c_{2\beta'}) t_Z }{2s_{2\bet'}}$& $ -\fr{c_Wt_Z }{2t_{\bet'}}$\\
\hline
$N_2$&$\fr 1 2 - \fr{(\Delta^2_{\mu}+\Delta^2_{\tau})c_W t_Z }{t_{2\bet'}}$&$\fr 1 2 +\fr{c_W t_Z t_{\bet'}}{2} $&$1+ \fr{c_W\left[s^2_{\beta'}-c^2_{\beta'}(\Delta^2_{\mu}+\Delta^2_{\tau})\right] t_Z }{s_{2\bet'}}$& $ -\fr{(\Delta^2_{\mu}+\Delta^2_{\tau}) c_Wt_Z }{2t_{\bet'}}$\\
\hline
$E_1$&$ -\fr 1 2 +  s^2_W +\fr{c_W t_Z}{ t_{2\bet'}} $&$ -\fr 1 2 +  s^2_W -\fr{c_W t_Z t_{\bet'}}{2} $&$  - 1 +  2 s^2_W +\frac{ (1-2t^2_{\beta'})c_W t_Z}{2 t_{\bet'}}$& $\frac{ c_W t_Z}{2 t_{\bet'}}$\\
\hline
$E_2$&$ -\fr 1 2 +  s^2_W -\fr{\left(\rho^2_{\mu\tau}-c^2_{\beta'}\right) c_W t_Z}{s_{2\bet'}} $&$ -\fr 1 2 +  s^2_W -\fr{c_W t_Z t_{\bet'}}{2} $&$  - 1 +  2 s^2_W +\frac{ (\rho^2_{\mu\tau}-c_{2\beta'})c_W t_Z}{s_{2\bet'}}$& $\frac{ (\Delta^2_{\mu}+\Delta^2_{\tau})c_W t_Z}{2 t_{\bet'}}$\\
\hline
$\overline{e}E_1,\;\overline{\nu_e}N_1$&$ \pm \fr{c_W t_Z }{2}$&$0$&$ \pm \fr{c_W t_Z }{2}$& $ \pm \fr{c_W t_Z }{2}$\\
\hline
$\overline{\nu_\mu}N_2,\;\overline{\nu_\tau}N_2 $&$ - \fr{\Delta_{\mu,\tau}\rho_{\mu\tau}c_W t_Z }{2s_{\beta'}}$&$0$&$ - \fr{\Delta_{\mu,\tau}\rho_{\mu\tau}c_W t_Z }{2s_{\beta'}}$& $ - \fr{\Delta_{\mu,\tau}\rho_{\mu\tau}c_W t_Z }{2s_{\beta'}}$\\
 \hline
 $\overline{\mu}E_2,\;\overline{\tau}E_2 $&$  \fr{\Delta_{\mu,\tau}\rho_{\mu\tau}c_W t_Z }{2s_{\beta'}}$&$0$&$  \fr{\Delta_{\mu,\tau}\rho_{\mu\tau}c_W t_Z }{2s_{\beta'}}$& $ \fr{\Delta_{\mu,\tau}\rho_{\mu\tau}c_W t_Z }{2s_{\beta'}}$\\
 \hline
$U_1$&$\fr 1 2 - \fr 2 3 s^2_W -\fr{c_W t_Z}{2 t_{2\bet'}} $&$\fr 1 2 - \fr 2 3 s^2_W +\fr{c_W t_Z t_{\bet'}}{2} $& $ 1 - \fr 4 3 s^2_W   +\frac{ (1-2t^2_{\beta'})c_W t_Z}{2 t_{\bet'}}$& $-\frac{ c_W t_Z}{2 t_{\bet'}}$\\
\hline
$U_2$&$\fr 1 2 - \fr 2 3 s^2_W +\fr{(\rho^2_{sb}-c^2_{\beta'})c_W t_Z }{2s_{\beta'}c_{\beta'}} $&$\fr 1 2 - \fr 2 3 s^2_W +\fr{c_W t_Z t_{\bet'}}{2} $& $ 1 - \fr 4 3 s^2_W +  \fr{(\rho^2_{sb}-c_{2\beta'})c_W t_Z }{2s_{\beta'}c_{\beta'}}$&$ -\fr{(\Delta^2_{s}+\Delta^2_{b}) c_Wt_Z }{2t_{\bet'}}$\\
 \hline
$D_1$ &$-\fr 1 2 + \fr 1 3 s^2_W +\fr{c_W t_Z}{ t_{2\bet'}} $& $-\fr 1 2 + \fr 1 3 s^2_W -\fr{c_W t_Z t_{\bet'}}{2} $ & $ - 1 + \fr 2 3 s^2_W  + \frac{ (1-2t^2_{\beta'})c_W t_Z}{2 t_{\bet'}}$& $\frac{ c_W t_Z}{2 t_{\bet'}}$\\
\hline
$D_2$ &$-\fr 1 2 + \fr 1 3 s^2_W -\fr{(\rho^2_{sb}-c^2_{\beta'})c_W t_Z }{2s_{\beta'}c_{\beta'}} $& $-\fr 1 2 + \fr 1 3 s^2_W -\fr{c_W t_Z t_{\bet'}}{2} $ & $ - 1 + \fr 2 3 s^2_W  - \fr{(\rho^2_{sb}-c_{2\beta'})c_W t_Z }{2s_{\beta'}c_{\beta'}} $&$ \fr{(\Delta^2_{s}+\Delta^2_{b}) c_Wt_Z }{2t_{\bet'}}$\\
\hline
$\overline{d}D_1,\;\overline{u}U_1$&$ \pm \fr{c_W t_Z }{2}$&$0$&$ \pm \fr{c_W t_Z }{2}$& $ \pm \fr{c_W t_Z }{2}$\\
\hline
$\overline{c}U_2,\;\overline{t}U_2 $&$ - \fr{\Delta_{s,b}\rho_{s,b}c_W t_Z }{2s_{\beta'}}$&$-\frac{\Delta_{s,b}m_{c,t}c_{\beta'}\epsilon}{2v\rho_{sb}s_{\beta}\widetilde{M}_{Q_2}}$&& \\
\hline
$\overline{s}D_2,\;\overline{b}D_2 $&$ \fr{\Delta_{s,b}\rho_{s,b}c_W t_Z }{2s_{\beta'}}$&$\frac{\Delta_{s,b}m_{s,b}c_{\beta'}\epsilon}{2v\rho_{sb}s_{\beta}\widetilde{M}_{Q_2}}$&& \\
\hline
\end{tabular}
\end{table}
Here we keep only significant contributions to $g_R$, in which they contain both factors of heavy masses and $\epsilon$, as shown in the two last lines of Table \ref{tZff}.
We can see that although new fermions are all vector-like in the flavor bases,
they are not vector-like in the mass bases because they are mixed with the chiral $SU(2)_2$ leptons
through Yukawa interactions~(\ref{g221e59}).

In contrast to~\cite{g221}, in our work the neutral currents are written in the basis
of \emph{physical} neutral gauge bosons, SM $Z$ and extra $Z'$,  from which their decays  can easily be  studied.
\subsection{Charged currents}
The Lagrangian of charged currents is
\bea
\mathcal{L}_{CC}
& = & \fr{1}{\sqrt{2}}\left\{ {\bar \nu}_L\ga^\mu g_2 W_{2 \mu}^+ l_L  + {\bar N}_{L,R}\ga^\mu g_1 W_{1 \mu}^+ E_{L,R}\right.
\crn & + & \left. {\bar u}_L\ga^\mu g_2 W_{2 \mu}^+ d_L  + {\bar U}_{L,R}\ga^\mu g_1 W_{1 \mu}^+ D_{L,R} \right\}+ \mathrm{H.c.},
\label{g221e48}
\eea
In the physical  states of charged gauge bosons, it is
\bea
\mathcal{L}_{CC} & = & \fr{g c_{\xi_W}}{\sqrt{2}}\left(1- \frac{t_{\xi_W}}{t_{\beta'}}\right)
W_{\mu}^+\left({\bar \nu}_L\ga^\mu  l_L  + {\bar u}_L\ga^\mu  d_L\right)\crn
&+& \fr{g c_{\xi_W}}{\sqrt{2}}\left(1- \frac{t_{\xi_W}}{t_{\beta'}}\right)  W_{\mu}^+(
{\bar N}\ga^\mu  E  + {\bar U}\ga^\mu  D )\crn
& - & \fr{g c_{\xi_W}}{\sqrt{2}}\left( \frac{1}{t_{\beta'}} + t_{\xi_W} \right)
W_{\mu}^{\prime +}\left({\bar \nu}_L\ga^\mu  l_L  + {\bar u}_L\ga^\mu  d_L\right)
\label{g221e54} \\
&+&  \fr{g c_{\xi_W}}{\sqrt{2}}\left( \frac{1}{t_{\beta'}} + t_{\xi_W} \right)
W_{\mu}^{\prime +}({\bar N}\ga^\mu  E  + {\bar U}\ga^\mu  D )
+ \mathrm{ H.c}.
\label{g221e56}
\eea
If the $W$ boson part of the Lagrangian is written as $ \mathcal{L} =  \fr{g c_{\xi_W}}{\sqrt{2}} W_{\mu}\overline{f}\gamma^{\mu}
\left(g_LP_L+g_RP_R\right)f' +\mathrm{H.c.}$,
the couplings of $W$ boson with physical fermions are shown in Table~\ref{tWff}.
\begin{table}[h]
  \centering
  \caption{Couplings of $W$ boson with fermions}\label{tWff}
  \begin{tabular}{|c|c|c|}
    \hline
    $f$ & $g_L$ & $g_R$ \\
      \hline
      $\overline{\nu_\ell}\ell$, $\ell=e,\mu,\tau$  &  $1-\Delta^2_{\ell}t_{\xi_W}$& $0$  \\
        \hline
      $\overline{u_i}d_i$, $i=1,2,3$  &  $1-\Delta^2_{d_i}t_{\xi_W}$& $0$  \\
      \hline
     $\overline{N_1}E_1$,  $\overline{U_1}D_1$ & $1-\frac{2t_{\xi_W}}{t_{2\beta'}}$ & $1+t_{\xi_W}t_{\beta'}$\\
      \hline
      $\overline{N_2}E_2$, $\overline{U_2}D_2$& $1+\rho^2_{\mu\tau,sb}t_{\xi_W}$ & $1+\frac{t_{\xi_W}}{t_{\beta'}}$\\
      \hline
      $\overline{\nu_e}E_1$,  $\overline{u}D_1$,  $\overline{N_1}e$, $\overline{U_1}d$& $-t_{\xi_W}$ & $0$\\
      \hline
      $\overline{U_2}s$, $\overline{U_2}b$& $-\frac{\rho_{sb}\Delta_{s,b}t_{\xi_W}}{s_{\beta'}}$ & $-\frac{c_{\beta'}\Delta_{s,b}}{\rho_{sb}s_{\beta}\widetilde{M}_{Q_2}}\times \frac{m_{s,b}}{v}\times \epsilon$\\
      \hline
    $\overline{c}D_2$, $\overline{t}D_2$& $-\frac{\rho_{sb}\Delta_{s,b}t_{\xi_W}}{s_{\beta'}}$ & $-\frac{c_{\beta'}\Delta_{s,b}}{\rho_{sb}s_{\beta}\widetilde{M}_{Q_2}}\times \frac{m_{c,t}}{v}\times \epsilon$\\
      \hline
  \end{tabular}
  \end{table}
New-physics interactions are in~(\ref{g221e54}). Within the experimental data on the $W$ decay width,
ones can get constraints on the mixing angles. That was discussed in detail in Ref.~\cite{g221}.
 \section{Higgs sector}
 \label{Higgssector}
From
\be
\Phi = \fr{1}{\sqrt{2}} \left(%
\begin{array}{cc}
\Phi^0 & \Phi^+ \\
-\Phi^-
& \tilde{\Phi}^0\\
\end{array}\,
\right)  \rightarrow  \Phi^\dag = \fr{1}{\sqrt{2}} \left(%
\begin{array}{cc}
\tilde{\Phi}^0 & - \Phi^+ \\
\Phi^-
& \Phi^0\\
\end{array}\,
\right),
\label{g221e64}
\ee
the potential is given as
\bea
V & = & \mu_\phi^2 \phi^\dag \phi +  \mu_{\phi '}^2 \phi'^\dag \phi' + \mu_\Phi^2 {\rm Tr}(\Phi^\dag \Phi)
+ \frac{\la_1}{2}  (\phi^\dag \phi)^2 +  \frac{\la_2}{2}   (\phi'^\dag \phi')^2
+  \frac{\la_3}{2}  [{\rm Tr}(\Phi^\dag \Phi)]^2\crn
&+& \la_4 (\phi^\dag \phi)(\phi'^\dag \phi') + {\rm Tr}(\Phi^\dag \Phi) [\la_5 (\phi^\dag \phi) + \la_6 (\phi'^\dag \phi')] - \mu (\phi'^\dag \Phi  \phi + \mathrm{H. c.})\, .
\label{g221e66}
\eea
Because the $\mu$ parameter is proportional to the squared masses of the charged and CP-odd  Higgs bosons,
it must be  positive with the minus sign before it in
the potential~(\ref{g221e66}).

The neutral scalars are expanded as
\be
\va^0 = \fr{1}{\sqrt{2}} (v_\phi + S_\phi + i A_\phi) \, ,
\va'^0 = \fr{1}{\sqrt{2}} (v_{\phi '}  + S_{\phi'}  + i A_{\phi '}) \, ,
\phi^0 = \fr{1}{\sqrt{2}} (u+ S_\Phi + i A_\Phi)\, .
\label{g221e67}
\ee
At the tree level, the minimum conditions of the Higgs potential are similar to the ones in
Ref.~\cite{g221}, except the opposite signs of $\mu$.
\bea
\mu_{\phi}^2 +\frac{1}{2}\left(  \la_1 v^2_\phi + \la_4 v_{\phi '}^2 + \la_5 u^2
- \frac{u\mu }{t_{\beta}}\right) & = & 0\, ,\crn
\mu_{\phi '}^2 +  \frac{1}{2}\left(  \la_2 v^2_{\phi '}  + \la_4 v_{\phi }^2
+ \la_6 u^2 -u \mu t_{\beta} \right) & = & 0\, ,
\label{g221e668} \\
\mu_{\Phi}^2 +   \frac{1}{2}\left( \la_5 v^2_\phi + \la_6 v_{\phi '}^2 + \la_3 u^2
-\frac{\mu}{u}v_{\phi}v_{\phi'} \right) & = & 0\, .\nn
\eea
Based on the minimum conditions, the parameter $\mu^2_{\phi,\phi',\Phi}$ can be expressed
as a function of the Higgs-self couplings $u$, $v$ and $\beta$. Next, the masses, mass eigenstates, and couplings of
 Higgs bosons will be calculated by inserting
these functions into the Higgs potential~(\ref{g221e66}).
\subsection{Squared mass matrices of the Higgs bosons}
In the original bases of singly charged and CP-odd neutral Higgs bosons
$\phi^{\pm}=(\varphi^\pm,\; \varphi'^\pm,\;\Phi^\pm )^T$ and $ A=(A_\phi,\; A_{\phi'},\;A_{\Phi} )^T$,
the corresponding squared mass matrices are
\be
M^2_{h^\pm}= \frac{\mu}{2} \left(
              \begin{array}{ccc}
                \frac{uc_{\beta}}{s_{\beta}} &-u  &v c_{\beta} \\
                 & \frac{us_{\beta}}{c_{\beta}}  &-vs_{\beta}  \\
                 &  &   \frac{v^2s_{\beta}c_{\beta}}{u}\\
              \end{array}
            \right), \;  M^2_{A}= \frac{\mu}{2} \left(
              \begin{array}{ccc}
                \frac{uc_{\beta}}{s_{\beta}} &-u  &-v c_{\beta} \\
                 & \frac{us_{\beta}}{c_{\beta}}  &vs_{\beta}  \\
                 &  &   \frac{v^2s_{\beta}c_{\beta}}{u}\\
              \end{array}
            \right).
\label{orm2ch}
\ee
In the basis of  CP-even Higgs bosons $S=(S_\phi, S_{\phi'} , S_\Phi  )^T$ the squared mass matrix
 $M_S^2$ corresponding the mass term $ \fr 1 2 S^T  M_S^2 S$ is
\bea
M^2_S = \left(%
\begin{array}{ccc}
\frac{u\mu c_{\beta}}{2s_{\beta}}+ \la_1 v^2 s^2_{\beta}  &-\frac{1}{2}u\mu+ \la_4 v^2 s_{\beta}c_{\beta}
&-\frac{1}{2}v\mu c_{\beta}+  \la_5  u vs_{\beta} \\
  &\frac{u\mu s_{\beta}}{2c_{\beta}} +  \la_2 v^2 c^2_{\beta}  & -\frac{1}{2}v\mu s_{\beta}+ \la_6 u v c_{\beta}\\
 & &\frac{v^2 \mu s_{\beta}c_{\beta}}{2u}+ \la_3 u^2\\
\end{array}\,
\right).
\label{g221e69}
\eea
The above matrices are consistent with those given in \cite{g221} after using the
relations~(\ref{g221e668}).
\subsection{Physical spectrum of Higgs bosons and their couplings}
We will find the Higgs bosons masses in two steps. At the first step, where $v\rightarrow 0$, all the
three squared  mass matrices are diagonalized through the same transformation
\be
C_1 = \left(
            \begin{array}{ccc}
              s_{\beta} & c_{\beta} & 0 \\
              0&  0& 1 \\
              c_{\beta}&-s_{\beta} & 0 \\
            \end{array}
          \right).
 \label{bto211}
\ee
In the second step, it is easy to determine  rotations diagonalizing the  squared mass matrices of charged and CP-odd neutral Higgs bosons.  By defining the mixing angle $\zeta$ as
\be
\sin\zeta\equiv \frac{v s_{\beta}c_{\beta}}{\sqrt{u^2+(v s_{\beta}c_{\beta})^2}} ,
\hs \cos\zeta\equiv \frac{u}{\sqrt{u^2+(v s_{\beta}c_{\beta})^2}},
\label{zangle}
\ee
the total mixing matrices used to diagonalize mass matrices in  (\ref{orm2ch}) are
\be
C_{h^\pm} =\left(
\begin{array}{ccc}
	-c_{\beta}s_{\zeta} & s_{\beta}s_{\zeta} & c_{\zeta} \\
	s_{\beta} & c_{\beta} & 0 \\
	c_{\beta}c_{\zeta} & - s_{\beta}c_{\zeta} & s_{\zeta} \\
\end{array}
\right) \; ,
C_{A} =\left(
\begin{array}{ccc}
	c_{\beta}s_{\zeta} &- s_{\beta}s_{\zeta} & c_{\zeta} \\
	s_{\beta} & c_{\beta} & 0 \\
	c_{\beta}c_{\zeta} & -s_{\beta}c_{\zeta} & -s_{\zeta} \\
\end{array}
\right).
\label{Cha}
\ee
Mass eigenstates of the charged and CP-odd Higgs bosons, denoted as
$H^{\pm}=(G^\pm_{1},\; G^\pm_{2}, h^{\pm} )^T$ and  $H_A=(G_{Z_1},\;G_{Z_2},\;h_{a})^T$,
 relate with the original states  through the following equations:
\be
\phi^{\pm }= C^T_{h^\pm} H^{\pm }, \hs \mathrm{and}\; A= C^T_{A} H_A.
\label{rela1}
\ee
Two linear combinations of $ G^\pm_{1}$ and $ G^\pm_{2}$ are  Goldstone bosons eaten up by the $W'^\pm$ and $W^\pm$  gauge bosons. Similarly,  linear combinations of $G_{Z_1}$ and $G_{Z_2}$ are  eaten up by $Z$ and $Z'$.
  There are two physical charged
Higgs bosons $h^{\pm}$ and one physical CP-odd neutral Higgs boson $h_a$  with masses $m^2_{h^\pm}$
and $m^2_{A}$, respectively. They satisfy
$C_{h^\pm} M^2_{h^\pm} C_{h^\pm}^T=\mathrm{diag}(0,0,m^2_{h^\pm})$ and
$C_{A} M^2_{A} C_{A}^T=\mathrm{diag}(0,0,m^2_{A})$, where
\be
m^2_{h^\pm}= m^2_{A}= \frac{\mu\left( u^2+ v^2s^2_{\beta}c^2_{\beta}\right)}{2 us_{\beta}c_{\beta}}.
\label{m2hca}
\ee
Regarding the CP-even neutral Higgs bosons,
after the  rotation (\ref{bto211}), the squared mass matrix is $M'^2_S =C_1M^2_SC_1^T$, which is a $3\times3$ matrix with following elements:
\bea \left(M'^2_S\right)_{11}&=& v^2\left( \lambda_1 s^4_{\beta} +\lambda_2 c^4_{\beta} +2\lambda_4 s^2_{\beta}c^2_{\beta} \right), \crn
\left(M'^2_S\right)_{12}&=&\left(M'^2_S\right)_{21}= v\left[ u\left( \lambda_5 s^2_{\beta} +\lambda_6 c^2_{\beta}\right)-\mu s_{\beta}c_{\beta}\right], \crn
\left(M'^2_S\right)_{13}&=&\left(M'^2_S\right)_{31}= v^2 s_{\beta}c_{\beta}\left[ \left(\lambda_4-\lambda_1\right)s^2_{\beta} +\left(\lambda_2-\lambda_4\right)c^2_{\beta} \right], \crn
\left(M'^2_S\right)_{22}&=&\lambda_3 u^2+ \frac{\mu v^2 s_{\beta} c_{\beta}}{2u}, \crn
\left(M'^2_S\right)_{23}&=&\left(M'^2_S\right)_{32}=\frac{v}{2}\left[ \left(c^2_{\beta} -s^2_{\beta}\right) \mu +\left( \lambda_6-  \lambda_5 \right) u 2s_{\beta}c_{\beta} \right], \crn
\left(M'^2_S\right)_{33}&=& \frac{\mu u}{2s_{\beta}c_{\beta}}+ \left( \lambda_1+ \lambda_2- 2\lambda_4 \right) v^2 s^2_{\beta}c^2_{\beta}.
\label{Mp2S}
\eea
In general, $M'^2_S$ is complicated and it cannot
be diagonalized exactly. Instead, using the parameter $\epsilon \equiv v/u\ll 1$,  we will find approximate solutions for mass eigenvalues,  keep terms up to the order of the electroweak scale. This is reasonable because  the SM-like Higgs boson mass  was found to be $125$ GeV.  Approximate solutions was used earlier to find consistent masses of
the lightest CP-even neutral Higgs bosons in supersymmetric models~\cite{cpvenHiggs}.
The mixing matrix will also be  determined approximately, corresponding to the mass eigenvalues.

We start from finding the eigenvalues of the matrix $M'^2_S $
 by solving the equation $\mathrm{ Det}\left(M'^2_S-\lambda I_3 \right)=0$, where $\lambda$ is expanded as
$\lambda= u^2 (\lambda_0+\lambda_1 \epsilon^2)$ to keep it up to the order of electroweak scale $v$.
We assume that $\lambda_0,\lambda_1\sim \mathcal{O}(1)$.
Using  $v=u \epsilon$,  we  can write  $\mathrm{ Det}\left(M'^2_S-\lambda I_3 \right)=a_0 +a_1\epsilon^2+ \mathcal{O}(\epsilon^4)=0$ where $a_0=a_0 (\lambda_0)$
and $a_1=a_1(\lambda_0,\lambda_1) $.  We will consider only the two following equations:
\bea
a_0(\lambda_0) &=&  -\lambda_0\left(\lambda_0-\frac{\mu u}{2s_{\beta}c_{\beta}} \right)
\left(\lambda_0- \lambda_3 u^2\right),
 \crn
a_1(\lambda_0,\lambda_1) \left.\frac{}{}\right|_{\lambda_0=0}
&\sim&u^2\left[ \left(  \lambda_1 s^4_{\beta} +\lambda_2 c^4_{\beta} +2\lambda_4 s^2_{\beta}c^2_{\beta}\right)
- \frac{1}{\lambda_3}\left( \lambda_5 s^2_{\beta} +\lambda_6 c^2_{\beta}
- \frac{\mu}{u} s_{\beta}c_{\beta} \right)^2 \right].
\label{fai}
\eea
The first equation in (\ref{fai}) shows that the largest contributions to Higgs masses
are the solutions of $a_0(\lambda_0)=0$, giving one zero and two non-zero values,
$\lambda_0= \frac{\mu u}{2s_{\beta}c_{\beta}} $ and  $\lambda_0= \lambda_3 u^2$.
Hence there are two heavy CP-even neutral Higgs bosons with the corresponding masses
$ m^2_{h^0_2}= \lambda_3 u^2 + \mathcal{O}(v^2)$ and $  m^2_{h^0_3} = \frac{\mu u}{2s_{\beta}c_{\beta}} + \mathcal{O}(v^2)$, which equal  the largest contributions of the two last diagonal entries of  $M'^2_S $ shown in (\ref{Mp2S}). A light CP-even neutral Higgs boson corresponds to $\lambda_0=0$. Its  mass comes from the second equation of (\ref{fai}):
\bea
m^2_{h^0_1}
 \simeq  v^2 \left[ \left(  \lambda_1 s^4_{\beta} +\lambda_2 c^4_{\beta}
+2\lambda_4 s^2_{\beta}c^2_{\beta}\right) - \frac{1}{\lambda_3}\left( \lambda_5 s^2_{\beta}
+\lambda_6 c^2_{\beta} - \frac{\mu}{u} s_{\beta}c_{\beta} \right)^2\right].
\label{ligh0}
\eea
We stress that $m^2_{h^0_1}\neq\left(M'^2_S\right)_{11}$, i.e.  contributions
from non-diagonal entries of $M'^2_S $ to $m^2_{h^0_1}$ cannot be ignored.  The mixing matrix in this case  can be found based on a mixing angle defined by  $c_h\equiv \cos\xi_h$ and $s_h\equiv \sin\xi_h$ satisfying
\be
t_{2h}= \tan 2\xi_h=-\frac{2\left(M'^2_S\right)_{12 }}{\left(M'^2_S\right)_{22 }
-\left(M'^2_S\right)_{11 }}\sim \fr v u=\epsilon.
\label{t2ch}
\ee
It can be checked that after this rotation the light Higgs boson mass is  consistent with ~(\ref{ligh0}). Therefore, the mixing matrix relating two original  and physical bases $S$ and  $ H_0=( h^0_1,h^0_2,h^0_3)^T $   is $S=C_{h}^T H_0$,
where
\be  C_{h}\simeq C_1 C_{h2}= \left(
                \begin{array}{ccc}
                 s_{\beta} c_h &c_{\beta} c_h & s_h \\
                  -s_{\beta}s_h &-c_{\beta} s_h & c_h \\
                  c_{\beta} & - s_{\beta} & 0 \\
                \end{array}
              \right).
\label{mixngCh}
\ee
The light Higgs boson $h^0_1$  is identified with the SM-like Higgs
boson found by the LHC. The recent experimental data shows that the SM predictions perfectly
agree with the observation within 1 $sigma$~\cite{Hdata}. Hence, couplings of $h^0_1$ with other
SM particles must be consistent with this data. The relevant couplings of $h^0_1$ are shown in Table \ref{h01SM},
\begin{table}[h]
  \centering
  \begin{tabular}{|c|c|}
  \hline
 Vertex & Coupling \\
 \hline
   $h^0_1 \overline{f_i}f_i$ &  $-\frac{c_h m_{e_i}}{v}$  \\
    \hline
   $h^0_1\overline{F_1}F_1$,  & $ -s_{h}c^2_{\beta'}\widetilde{M}_{L_1,Q_1}$ \\
 \hline
 $h^0_1\overline{E_2}E_2$ & $-s_{h}c^2_{\beta'}(\Delta^2_{\mu}+\Delta^2_{\tau})\widetilde{M}_{L_2}$  \\
  \hline
      $h^0_1\overline{U_2}U_2$, $h^0_1\overline{D_2}D_2$ & $-s_{h}c^2_{\beta'}(\Delta^2_{s}+\Delta^2_{b})\widetilde{M}_{Q_2}-\frac{c_hc^2_{\beta'}(\Delta^2_{s}m^2_{c,s} +\Delta^2_{b}m^2_{t,b})\times \epsilon}{s_{\beta}v^2\widetilde{M}_{Q_2}\rho^2_{sb}}$   \\
  \hline
  $h^0_1 \overline{e}E_1$,  $h^0_1 \overline{q_1}Q_1$ & $ -s_h s_{\beta'}c_{\beta'}\widetilde{M}_{L_1,Q_1}P_L $  \\
  \hline
   $h^0_1 \overline{\mu}E_2$,   $h^0_1 \overline{\tau}E_2$ &  $ -\Delta_{\mu,\tau}c_{\beta'} \left[s_h \widetilde{M}_{L_2}\rho_{\mu\tau}P_L+ \frac{c_h m_{\mu,\tau}}{v\rho_{\mu\tau}}P_R\right]$ \\
  \hline
     $h^0_1 \overline{q_{2}}Q_2$, $h^0_1 \overline{q_{3}}Q_2$  &  $ -\Delta_{s,b}c_{\beta'} \left[s_h \widetilde{M}_{Q_2}\rho_{sb}P_L+   \frac{ c_h m_{q_{2,3}}}{v\rho_{sb}}\left(P_R+ \frac{ m_{q_{2,3}}}{v \widetilde{M}_{Q_2}s_{\beta}}\epsilon P_L \right)\right]$\\
   \hline
  $h^0_1Z_{\mu}Z_{\nu}$&$\frac{g^{\mu\nu}g^2 v }{2c_W^2}c_h$\\
  \hline
$h^0_1W^+_{\mu}W^-_{\mu}$& $\frac{g^{\mu\nu}g^2 v}{2} c_h $\\
\hline
$h^0_1W'^+_{\mu}W'^-_{\mu}$&$\frac{g^{\mu\nu} g^2\left[4us_h+c_hv\left(1-2c_{2\beta}c_{2\beta'}+c_{2\beta'}^2\right)\right]}{8c_{\beta'}^2s_{\beta'}^2} $\\ \hline
$ h^0_1h_1^+h_1^- $ &
$-s_hu \left[\frac{\mu}{u} c_\beta s_\beta-\left(\lambda_5c^2_{\beta}+\lambda_6 s^2_{\beta}\right)\right]-c_h v \left[\frac{\mu}{u} c_\beta s_\beta+ (\lambda_1+\lambda_2)s^2_{\beta}c^2_{\beta}
+\lambda_4(s^4_{\beta}+c^4_{\beta})\right] $ \\
\hline
$ h^0_1h_2^+h_2^- $ & $-us_h\lambda'_3 -vc_h\left( \lambda'_2c^2_{\beta}+\lambda'_1s^2_{\beta}\right)$ \\ \hline
 \end{tabular}
\caption{$h^0_1$ couplings , where  $f_i=e_i,u_i,d_i$; $q_i=u_i,d_i$  ($i=1,2,3)$; $q_1=u,d$; $Q_1=U_1,D_1$;
$F_1=E_1,U_1,D_1$.}
\label{h01SM}
\end{table}
including couplings with
 $h^\pm_2$  needed to generate active neutrino masses. We can see easily that all couplings  with the SM-like particles
are different from the SM predictions by a common  factor  $c_h$.
So, $|c_h|$ should be close to unity, i.e., $|s_h|$ should be small.  Its upper bound can be found as follows.
 Consider the $h^0_1$ productions at LHC, new heavy quarks can play the roles of the top quark
in gluon-gluon fusion mechanism, where their couplings are proportional to $s_h$ or $\epsilon$. A significant contribution related to $\epsilon$  may come from the quarks $U_2$ where the couplings contain a
factor $(\Delta_b m_t)^2\epsilon/v^2$. But the constraint from~\cite{g221} gives  $(\Delta_b m_t)^2\epsilon/v^2\sim 10^{-4}\epsilon$, which is suppressed. Now, the production of $h^0_1$ through gluon-gluon fusion at lowest order  is  \cite{higgsprodecay}
\be
\sigma^0_{h^0_1}=\frac{G_F \alpha_s^2 }{288\sqrt{2}\pi}\left|\frac{3}{4}\sum_{q}\frac{g_{hqq}v}{m_q}
A_{1/2}(t_{q})\right|^2,
\label{ggH}
\ee
where $t_{q}=\frac{m^2_{h^0_1}}{4 m_q^2}$, $m_q$ is a quark mass; $g_{hqq}=c_h \frac{m_q}{v}$,
$s_h c^2_{\beta'}\widetilde{M}_{Q_1}$, and $s_h c^2_{\beta'}\left(\Delta^2_s +\Delta^2_b\right)\widetilde{M}_{Q_2}$
for SM-like quarks; new quarks $U_1,D_1$; and new quarks $U_2,D_2$, respectively.
The form factor $A_{1/2}(t)$ is determined as
\be
A_{1/2}(t)=2\left[t+(t-1)f(t)\right]t^{-2} \label{a1p2},
\ee
where
\[
f(t)\equiv\left\lbrace
\begin{matrix}
	\arcsin^2\left(\sqrt{t}\right)& \ \mathrm{for}\ t\leq  1\\
	-\frac{1}{4}\left(-i\pi+\ln\left[\frac{1+\sqrt{1-t^{-1}}}{1-\sqrt{1-t^{-1}}}\right]\right)^2 & \ \mathrm{for}\ t> 1.\\
\end{matrix}\right.
\]
Using  $m_{Q_{1,2}}$,  $m_{Q_{1,2}}\simeq\widetilde{ M}_{Q_{1,2}}u$, as given in  \cite{g221},
Eq.~(\ref{ggH}) is written as
\be
\sigma^0_{h^0_1}=\frac{G_F \alpha_s^2 }{288\sqrt{2}\pi}\left|\frac{3}{4} c_h A_{1/2}(t_{0})
+ \frac{3}{2} c^2_{\beta'}s_h \epsilon \left[ A_{1/2}(t_{1})+(\Delta_{s}^2+\Delta^2_b)A_{1/2}(t_{2})\right]\right|^2,
\label{ggh2}
\ee
where $t_0=\frac{m_{h^0_1}}{4m^2_t}$ and  $t_{1,2}=\frac{m_{h^0_1}}{4m^2_{Q_{1,2}}}$.
The condition  $m_t, m_{Q_{1,2}}> m_{h^0_1}=125.09$ GeV gives the limit $A_{1/2}(t)\rightarrow 4/3$
for all $t=0,1,2$.
 The respective signal strength of Higgs production is
\be
\mu_{ggF}=\fr{\sigma(pp\rightarrow h^0_1)_{\mathrm{221}}}{\sigma(pp\rightarrow h)_{\mathrm{SM}}}
=\left| c_h +2c^2_{\beta'}s_h\epsilon \left[ 1+\Delta_{s}^2+\Delta^2_b\right]\right|^2,
\label{signalggh}
\ee
where we follow the notations of signal strengths defined in~\cite{Hdata}.
Similarly, the partial decay width of the channel $h^0_1\rightarrow gg$ is determined as
\be
\Gamma(h^0_1\rightarrow gg)_{221}= \mu^{hgg}\times \Gamma(h\rightarrow gg)_{\mathrm{SM}},
\label{dhgg}
\ee
where $\mu^{hgg}=\mu_{ggF}$. Because $s_h \epsilon=\mathcal{O}(\epsilon^2)$ and the branching
ratio of this decay is smaller than $9\%$, we will use the naive approximation
$\mu_{ggF}\simeq c^2_h$ to find a lower bound of $|c_h|$.

For all remaining decay channels of the SM-like Higgs boson into SM particles, the tree-level
couplings are always different from the SM prediction the factor $c_h$, therefore $\mu^f=c_h^2$ for
 all main decays $f=f\bar{f},\;WW^*,\;ZZ^*$.  The global signal strength defined in~\cite{Hdata} can be formulated  approximately  as follows
\be
\mu^f_i(\mathrm{global})=\mu_i\times\mu^f\simeq |c_h|^4=1.09\pm 0.11,
\rightarrow 0.98\leq |c_h|^4\leq1.\nn
\label{globalmu}
\ee

This gives the constraint
\be
0.995\leq|c_h|\leq 1,\hs  \mathrm{and}\; |s_h|\leq0.10.
\label{cshconstrain}
\ee
If we use the constraints of $\Delta_{b,s}$ and $c_{\beta'}$ given in \cite{g221},
the values of $|s_h|$ satisfying  (\ref{cshconstrain}) are reasonable for the approximation
we have just discussed. In addition, the constraint of $s_h$ results in small couplings of
$h^0_1$ with the heavy fermions and $W'$ gauge boson, giving their  suppressed contributions to the decay rate  $h^0_1\rightarrow \gamma\gamma$. Couplings $h^0_1h^{\pm}_{1,2}h^{\pm}_{1,2}$  depend on many unknown Higgs-self couplings,
implying that   charged Higgs masses are not constrained from the experimental data of the  decay  $h^0_1\rightarrow \gamma\gamma$,
so we will not consider this decay further.
\subsection{Singly charged Higgs bosons with additional $\delta^\pm$}
\label{modhiggs}
In this section we consider the model including new singly charged Higgs bosons $\delta^\pm$ discussed in the neutral lepton sector. Apart from the second term in (\ref{Lnumass}),
the Higgs potential has new  terms,
\bea
\Delta V_h &=& \mu^2_4(\delta^+\delta^-)+ \lambda'_{0}(\delta^+\delta^-)^2
+ (\delta^+\delta^-)\left[ \lambda'_1 \mathrm{Tr}(\Phi^\dagger\Phi) + \lambda'_2\phi^\dagger\phi + \lambda'_3\phi'^\dagger\phi' \right] .
\label{aVh}
\eea
The appearance of $\delta^\pm$ does not change the allowed regions of parameters discussed in~\cite{g221}. Also, the results derived for  Higgs bosons are unchanged, except  the singly
charged Higgs sector. In the basis $(\varphi^\pm,\; \varphi'^\pm,\Phi^\pm,\delta^\pm)^T$, the squared mass matrix is denoted as $\mathcal{M}^2_{h^\pm}$.   We can find a matrix $C'_{h^\pm}$  so that  $\mathcal{M}'^2_{h^\pm}=C'_{h^\pm}\mathcal{M}'^2_{h^\pm}C'_{h^\pm T}$ has only the following non-zero elements: $\left(\mathcal{M}'^2_{h^\pm}\right)_{34}=\left(\mathcal{M}'^2_{h^\pm}\right)_{43}=\frac{1}{2}\lambda_{\delta}  v\sqrt{u^2+(vs_{\beta}s_{\beta})^2} $, $\left(\mathcal{M}'^2_{h^\pm}\right)_{33}=\frac{\mu\left[u^2+(vs_{\beta}c_{\beta})^2\right]}{2us_{\beta}c_{\beta}}$ and $\left(\mathcal{M}'^2_{h^\pm}\right)_{44}=\mu^2_4+\frac{1}{2}v^2\left( \lambda'_1  s^2_{\beta}+ \lambda'_2 c^2_{\beta} \right) + \frac{1}{2}\lambda'_3  u^2$.
  This matrix is diagonalized by  a transformation relating with a mixing angle $\xi$ satisfying
\be
t_{2\xi}\equiv \tan 2\xi=-\frac{2  \left(\mathcal{M}'^2_{h^\pm}\right)_{34}}{\left(\mathcal{M}'^2_{h^\pm}\right)_{44}
-\left(\mathcal{M}'^2_{h^\pm}\right)_{33}}.
\label{xia}
\ee
Then the total transformation  can be found to be
\[ C_{\pm}= \left(
                  \begin{array}{cccc}
                  -c_{\beta}s_{\zeta}& s_{\beta} s_{\zeta}&c_{\zeta} & 0\\
                   s_{\beta} &c_{\beta} &0 &0  \\
                  c_{\beta}c_{\zeta}c_{\xi} & -s_{\beta}c_{\zeta}c_{\xi} &s_{\zeta}c_{\xi} & -s_{\xi} \\
                   c_{\beta}c_{\zeta}s_{\xi} &  -s_{\beta}c_{\zeta}s_{\xi} & -s_{\zeta}s_{\xi} & c_{\xi}\\
                  \end{array}
                \right),
\]
which  changes the original basis  into the mass eigenstate basis $( G^\pm_1, \; G^\pm_2,\;h^\pm_1,\; h^\pm_2)^T$, namely
$(\varphi^\pm,\; \varphi'^\pm,\; \Phi^\pm,\delta^\pm)^T=C^T_{\pm}( G^\pm_1,\; G^\pm_2,\; h^\pm_1,\; h^\pm_2)^T$.
We note that the Goldstone bosons $G^\pm_{1,2}$ defined in~(\ref{rela1})  are not affected by the presence of $\delta^\pm$.

The masses $m^2_{h^\pm_{1,2}}$ of  $h^{\pm}_{1,2}$ are solutions of the equation
$
\left[  x-\left(\mathcal{M}'^2_{h^\pm}\right)_{33} \right] \left[  x- \left(\mathcal{M}'^2_{h^\pm}\right)_{44} \right] +  \left(\mathcal{M}'^2_{h^\pm}\right)_{34}^2=0
$.
If $\left(\mathcal{M}'^2_{h^\pm}\right)_{34}\ll \left(\mathcal{M}'^2_{h^\pm}\right)_{33},\left(\mathcal{M}'^2_{h^\pm}\right)_{44}$,  we have $h^\pm_1\equiv h^\pm$
given in~(\ref{rela1}) and $h^\pm_2\equiv \delta^\pm$ which is  used for simple approximations because Eq.~(\ref{xia}) means $t_{2\xi}\sim \epsilon\ll1$.  The relevant couplings of the charged Higgs bosons to fermions
are collected in Table~\ref{chffcoupling}.
\begin{table}[h]
  \centering
   \begin{tabular}{|c|c|}
     \hline
   Vertex   & Coupling \\
   \hline
    $h^+_1 \overline{\nu_{L_i}} e_{R_i}$ , $i=e,\mu,\tau$ & $-\frac{\sqrt{2}c_{\zeta}m_{e_i}}{t_{\beta}v}\times c_{\xi}$\\
     \hline
    $h^+_1 \overline{(\nu_{L_\mu})^c} e_L$,    $h^+_1 \overline{(\nu_{L_e})^c} \mu_L$ & $\mp 2s_{\beta'}s_{\xi}\left[ f_{12}-\frac{c^2_{\beta'}\Delta_{\mu}\left(f_{12}\Delta_\mu +f_{13}\Delta_\tau\right)}{1+\rho_{\mu\tau}}\right]$\\
        \hline
     $h^+_1 \overline{(\nu_{L_\tau})^c} e_L$, $h^+_1 \overline{(\nu_{L_e})^c} \tau_L$ & $\mp 2s_{\beta'}s_{\xi}\left[ f_{13}-\frac{c^2_{\beta'}\Delta_{\tau}\left(f_{12}\Delta_\mu +f_{13}\Delta_\tau\right)}{1+ \rho_{\mu\tau}}\right]$\\
       \hline
     $h^+_1 \overline{(\nu_{L_\tau})^c} \mu_L$, $h^+_1\overline{(\nu_{L_\mu})^c} \tau_L$ & $\mp 2s_{\xi}f_{23}\rho_{\mu\tau} $\\
        \hline
       $h^+_1 \overline{N_1} e$ & $-\sqrt{2} s_{\beta'} c_{\beta'} s_{\zeta}c_{\xi}\widetilde{M}_{L_1}P_L-\frac{\sqrt{2}c_{\zeta}c_{\xi}m_e}{t_{\beta}t_{\beta'}v}P_R$\\
        \hline
       $h^+_1 \overline{N_2} \mu$,  $h^+_1 \overline{N_2}\tau$ & $-\sqrt{2}c_{\xi}c_{\beta'}\Delta_{\mu,\tau}\left[\frac{c_{\zeta}m_{\mu,\tau}}{t_{\beta}v \rho_{\mu\tau}}P_R + s_{\zeta} \rho_{\mu\tau} \widetilde{M}_{L_2}P_L \right]$\\
       \hline
       $h^+_1 \overline{(N_{L_2})^c} e_L$ & $-2s_{\beta'}c_{\beta'}s_{\xi}\left( \Delta_{\mu}f_{12} +\Delta_{\tau}f_{13}\right)$\\
        \hline
       $h^+_1 \overline{(N_{L_1})^c}\mu_L$,   $h^+_1 \overline{(N_{L_1})^c}\tau_L$&  $ 2c_{\beta'}s_{\xi}\left[ f_{12,13}-\frac{c^2_{\beta'}\Delta_{\mu,\tau}\left(f_{12}\Delta_\mu +f_{13}\Delta_\tau\right)}{1+\rho_{\mu\tau}}\right]$\\
       \hline
       $h^+_1 \overline{(N_{L_2})^c}\mu_L$,   $h^+_1 \overline{(N_{L_2})^c}\tau_L$&  $ \mp 2c_{\beta'}s_{\xi}\Delta_{\tau,\mu}f_{23}$\\
       \hline
   $h_1^+ \overline{u}_id_i$, $i=1,2,3$    & $\frac{\sqrt{2}c_{\xi}c_{\zeta}}{vt_{\beta}}\left( m_{u_i} P_L-m_{d_i} P_R\right)$ \\
   \hline
  $h^+_1 \overline{U_1}d_1$ & $-\sqrt{2} s_{\beta'} c_{\beta'} s_{\zeta}c_{\xi}\widetilde{M}_{Q_1}P_L-\frac{\sqrt{2}c_{\zeta}c_{\xi}m_d}{t_{\beta}t_{\beta'}v}P_R$\\
    \hline
    $h_1^+ \overline{U}_2s$, $h_1^+ \overline{U}_2b$&   $-\sqrt{2}c_{\xi}s_{\beta'}\Delta_{s,b}\left[\frac{c_{\zeta}m_{s,b}}{t_{\beta}v \rho_{sb}}P_R + s_{\zeta} \rho_{sb} \widetilde{M}_{Q_2}P_L \right]$\\
     \hline
     $h_1^+ \overline{u}D_1$ & $\frac{\sqrt{2}c_{\xi}c_{\zeta}m_{u}}{vt_{\beta}t_{\beta'}} P_L$ \\
     \hline
   $h_1^+ \overline{c}D_2$, $h_1^+ \overline{t}D_2$    & $ \frac{\sqrt{2}c_{\xi}s_{\zeta}c_{\beta'}m_{c,t}\Delta_{s,b}}{ vt_{\beta}} P_L$ \\
    \hline
     $h_1^+ \overline{U_1}D_1$    & $- \sqrt{2}c_{\xi}s_{\zeta}c^2_{\beta'}\widetilde{M}_{Q_1} P_L$ \\
   \hline
   $h_1^+ \overline{U_2}D_2$ &  $- \sqrt{2}c_{\xi}c^2_{\beta'}\left[\frac{c_{\zeta}\epsilon\left[\Delta_b^2\left(-m_t^2P_L+m_b^2P_R\right) +\Delta_s^2\left(-m_c^2P_L+m_s^2P_R\right)\right]}{v^2s_{\beta}t_{\beta}\widetilde{M}^2_{Q_2} \rho^2_{sb}}+s_{\zeta}(\Delta^2_{s} +\Delta^2_b) \widetilde{M}_{Q_2} P_L\right]$\\
        \hline
\end{tabular}
\caption{Couplings of charged Higgs boson $h^+_1$. Couplings of $h^{\pm}_2$ are only different from those of $h^\pm_1$ by replacements: $c_{\xi}h^\pm_1 \rightarrow s_{\xi} h^\pm_2$ and $s_{\xi} h^\pm_1\rightarrow -c_{\xi} h^\pm_2$. }
\label{chffcoupling}
\end{table}

Couplings of the charged Higgs bosons with gauge bosons are shown in Table~\ref{chboson}.
Only couplings of $h^\pm_1$ are shown because the couplings of $h^\pm_2$ can be derived by the following replacements:
$c_{\xi}h^\pm_1 \rightarrow s_{\xi} h^\pm_2$ and $s_{\xi} h^\pm_1\rightarrow -c_{\xi} h^\pm_2$.  We consider here only the case of  $\xi\rightarrow0$.
\begin{table}[h]
  \centering
  \begin{tabular}{|c|c|}
    \hline
  Vertex   & Coupling \\
     \hline
        $h^+_1 Z_{\mu}W^-_{\nu}$&  $-\frac{g^{\mu\nu} gm_Wc_{\xi}s^3_{\beta}c^3_{\beta}s_W^2c_{Z}s_Z\epsilon^2}{2 s_{\beta'}c_{\beta'}}$\\
     \hline
     $h^+_1 Z_{\mu}W'^-_{\nu}$&$-\frac{g^{\mu\nu}gm_Wc_{\xi}s_{\beta}c_{\beta}\left(s_W^2+c^2_Zc^2_W\right)}{c_W s_{\beta'}c_{\beta'}}$\\
       \hline
          $h^+_1 Z'_{\mu}W^-_{\nu}$&$-\frac{g^{\mu\nu}gm_Wc_{\xi}s_{\beta}c_{\beta}}{s_{\beta'}c_{\beta'}}$\\
       \hline
          $h^+_1 Z'_{\mu}W'^-_{\nu}$&$-\frac{g^{\mu\nu}gm_Wc_{\xi}s_{\beta}c_{\beta}c_{\zeta}s_{\zeta}s^2_W}{c_Ws_{\beta'}c_{\beta'}}$\\
       \hline
     $h^+_1 h^0_1W^-_{\mu}$& $\frac{gc_{\xi}s_Ws_{\beta}c_{\beta}}{2 s_{\beta'}c_{\beta'}}s_{Z}\left(c_h-s_h\epsilon\right) \left( p_0-p_+\right)^{\mu}$ \\
     \hline
     \end{tabular}
\caption{ Couplings of $h^+_1$ with  bosons. Momenta of $h^+_1$ and $h^0_1$ are $p_+$ and $p_0$, respectively.}
\label{chboson}
\end{table}

Some important properties of $h^\pm_1$ are as follows. Differences between couplings of $h^\pm_1$ and the SM-like Higgs bosons to normal fermions are $ g_{h^\pm_1\ell \nu_{\ell}}/g_{h^0_1\ell\ell}=\frac{\sqrt{2}c_{\xi}c_{\zeta}}{t_{\beta}}$ and $ g_{h^\pm_1u_i d_i}/g_{h^0_1u_iu_i}=-\frac{\sqrt{2}c_{\xi}c_{\zeta}}{t_{\beta}}\left( P_L-\frac{m_{u_i}}{m_{d_i}}P_R\right)$. From Table~\ref{chboson}, the couplings of  $h^\pm_1$ to the SM-like bosons, namely $h^\pm_1 Z W$ and $h^\pm_1 h^0_1 W$,
are extremely small because they contain factors $s_Z\epsilon^2\sim \mathcal{O}(\epsilon^4)$,  $s_Z\sim \mathcal{O}(\epsilon^2)$, and another mixing  smaller than $0.02$.
Other couplings to light fermions  are also small because $s_{\xi}\rightarrow0$. With  $c_{\zeta}, c_{\xi}\rightarrow 1$,  the main decays of $h^\pm_1$ into light particles are  $h^+_1\rightarrow t\overline{b}$. If  $m_{h^{\pm}_1}> m_{W'}, m_{Z'}$,
there will appear two additional large decay modes $h^+_1\rightarrow Z'W^+, ZW'^+$.

In contrast to $h^\pm_1$, the charged Higgs bosons $h^\pm_2$ only couple  strongly with leptons and
Higgs bosons. Therefore, the main decay modes are $ h^+_2\rightarrow \overline{(\nu_{e_i})^c}e_j$ with $e_{j}=e,\mu,\tau$. The main processes for $h^{\pm}_2$ production at colliders are
$\overline{f}f\rightarrow \gamma^*,h^{0*}_{1,2,3},h^*_a\rightarrow h^+_2h^-_2$, $f=e,u,d$.

We would like to compare the above singly charged Higgs bosons with the ones predicted by Zee models,  where the charged Higgs sector
 was investigated thoroughly in Ref.~\cite{chzee}. The equivalent
notations are $v_1, v_2, \tan{\beta}=v_2/v_1\leftrightarrow v_{\phi},v_{\phi'}, 1/t_{\beta}=v_{\phi'}/v_{\phi}$,
and  $\chi\leftrightarrow\xi$.  With  $\zeta=\mathcal{O}(\epsilon)\rightarrow 0$,
the  charged components of $\Phi$ are Goldstone bosons of $W'^\pm$. Meanwhile those of $\phi$
and $\phi'$ create  two
Goldstone bosons of $W^\pm$, and two other freedoms that mix with the singlet ones to generate physical states. Hence, the model under consideration and the Zee models predict very similar properties of the
 charged Higgs couplings to SM-like particles.
But the predictions for charged Higgs boson  production
 at colliders like the LHC are different, because of  the appearance of new particles, such as new heavy quarks and Higgs bosons, and   the constraints from  allowed regions of parameters indicated in~\cite{g221}. We will review these regions before discussing  the signal of new particles at colliders.
\section{Phenomenology}
\label{pheno}
\subsection{ Properties of masses and mixing parameters of new particles}
In this work, the  results of parameter constraints
reported in Ref.~\cite{g221} are still
valid. They will be used to discuss the Higgs phenomenology. These allowed regions are
\bea
\Delta_s &\in& [-1.16,-0.97],\hs \Delta_b \in [0.003,0.007], \hs \Delta_{\mu} \in [0.94, 0.99], \hs \Delta_{\tau} \in [0,0.11],
\crn
M_{Z'}&\in& [500, 1710]\; \mathrm{GeV}, \hs \frac{g}{s_{\beta'}} \in [1.2,3.5], \hs \zeta'\equiv(s^2_{\beta}-t^2_{\beta'}c^2_{\beta})
\in [0,0.02],
\label{allowg221}
\eea
where $g=\frac{2m_W}{v}\simeq 0.651$ is the   SM gauge coupling,
and $\zeta'$ satisfies $t_{\xi_W}=c_W t_{Z}\simeq c^3_{\beta'}s_{\beta'}\zeta'\epsilon^2$ \cite{g221}.
This gives the constraints
\be
s_{\beta'}\in [0.186,0.542],\hs t_{\beta}\in [0.24,0.654], \hs \mathrm{and}\;
0\leq t_{Z}=\frac{t_{\xi_W}}{c_W}<8\times 10^{-3}\epsilon^2.
\label{tzxw}
\ee
We can see that the allowed values of $\zeta$ give very small values of $t_{Z,\xi_W}$, even with large $\epsilon<1$.
For simplicity,  we will also use the following simple approximations:
\be
|\Delta_{\mu,s}|\simeq 1, \hs \Delta_{\tau,b},\frac{\Delta_{\tau,b} m_{\tau,t}}{v}\ll1,
\hs
\Delta^2_{\mu,s}+\Delta^2_{\tau,b} \simeq 1, \hs \rho_{\mu\tau,bs}\simeq s_{\beta'}.
\label{appro}
\ee
 The simple texture of $\lambda_\ell$ in  (\ref{laell1})   gives  $ M_{L_1}= u \widetilde{M}_{L_1} s_{\beta'}$ and
$M_{L_2}= u \widetilde{M}_{L_2}\rho_{\mu\tau}$.
 For  the quark sector,
$M_{Q_1}=u \widetilde{M}_{Q_1} s_{\beta'}$ and $M_{Q_2}= u \widetilde{M}_{Q_2}\rho_{sb}$. Now the masses of the heavy particles are
\bea m_{E_1}&=&M_{N_1} +  \frac{m^2_{e}(s_{\beta}-\frac{1}{2})}{ M_{N_1} s^2_{\beta}t^2_{\beta'}}, \hs
m_{U_1,D_1}= u\widetilde{M}_{Q_1} +\frac{m^2_{u,d}(s_{\beta}-\frac{1}{2})}{ u\widetilde{M}_{Q_1} s^2_{\beta}t^2_{\beta'}},
\crn
m_{E_2}&=&M_{N_2} - \frac{c^2_{\beta'}(\Delta^2_{\mu}m^2_{\mu}+ \Delta^2_{\tau}m^2_{\tau})(s_{\beta}
-\frac{1}{2})}{ M_{N_2} s^2_{\beta}\rho^2_{\mu\tau}},
\crn
m_{U_2,D_2}&=& u\widetilde{M}_{Q_2} -\frac{c^2_{\beta'}(\Delta^2_{s}m^2_{c,s}
+ \Delta^2_{b}m^2_{t,b})(s_{\beta}-\frac{1}{2})}{ u\widetilde{M}_{Q_2} s^2_{\beta}\rho^2_{sb}},
\crn
m_{W'},m_{Z'}&\simeq& \frac{g u}{s_{2\beta'}}+  \frac{s^2_{2\beta}+ (c_{2\beta}-c_{2\beta'})^2}{8} m_W \times \epsilon
+  \mathcal{O}(\epsilon^3),
\crn
m_{h^0_3},\;m_A,\; m_{h^\pm_1}&\simeq&  \sqrt{\frac{u\mu}{s_{2\beta}}} + v\times \mathcal{O}\left(\frac{vs_{2\beta}}{\sqrt{u\mu}}\right),
\label{Mf}
\eea
and the mixing parameters
\bea
s_{Z}&\simeq& \frac{s_{\xi_W}}{c_W}\simeq\frac{t_{2\xi_W}}{2c_W} \simeq
\frac{1}{4} s_{2\beta'} \left(c_{2\beta}-c_{2\beta'}\right)\epsilon^2 , \hs
\crn
s_{\zeta} &\simeq& s_{\beta}c_{\beta}\epsilon, \hs c_{\zeta} \simeq 1- \frac{1}{2} (s_{\beta}c_{\beta})^2\epsilon^2,
\crn
s_{h}&\simeq& \frac{\lambda_5 s^2_{\beta} +\lambda_6 c^2_{\beta} -\frac{\mu}{u}s_{\beta}c_{\beta}}{\lambda_3}\times\epsilon, \hs
c_h \simeq 1 -\frac{s^2_h}{2},
\crn
s_{\xi}&\simeq& \frac{\lambda_{\delta}}{2\mu^2_4/u^2+\frac{\mu}{(us_{\beta}c_{\beta})}+\lambda'_{3}}\times \epsilon, \hs
c_{\xi}\simeq 1-\frac{s^2_{\xi}}{2}.
\label{mixing}
\eea
Inserting formula of $s_h$ into Eq. (\ref{ligh0}), we have a simple expression for $m_{h^0_1}$ as follows:
\bea
m^2_{h^0_1}&\simeq&   v^2  \left(  \lambda_1 s^4_{\beta} +\lambda_2 c^4_{\beta} +2\lambda_4 s^2_{\beta}c^2_{\beta}\right)
- \lambda_3 u^2s_h^2.
\label{mligh0}
\eea
In addition, there are two other Higgs bosons with the masses
\be
m_{h^0_2}\simeq \sqrt{\lambda_3}u, \hs \mathrm{and}\;m^2_{h^\pm_2}\simeq \mu^2_{4}
+\frac{v^2}{2}\left( \lambda'_1s^2_{\beta}+ \lambda'_2c^2_{\beta}\right) + \frac{\lambda'_3}{2}u^2.
\label{nhmass}
\ee
The important property is that the model predicts several groups of new heavy particles having same spins
and degenerate masses,  therefore forbid many decay modes.
The new Yukawa couplings  generating heavy mass terms are always the same for both up
and down components of the $SU(2)_1$ fermion doublets in the same families.  Mass differences come only from the Yukawa terms of the electroweak sector~(\ref{g221e57}),
because of the large $V_{Q,L}$.  As shown in~(\ref{Mf}), a difference between
a pair of new up and down fermions is $\mathcal{O}(m_f)$.
 Hence, the top quark may give the largest difference if  $\Delta_{b}m_t<1.4$ is not considered. Hence, the mass differences are always smaller than $m_W$. Because of the kinetic condition and the fermion number conservation, a three-body decay of a new fermion must decay to at least a light fermion, namely  a  SM fermion plus a boson.  If a fermion is the lightest among the new particles, it  will decay only to a light fermion and a SM boson like $W$, $Z$ or $h^0_1$, leading to  large branching ratios,  which can be searched  by recent colliders.
\subsection{Searches for new fermions at colliders}
From the above discussion, if the new fermions are lighter than all new bosons, including $W'^\pm, Z', h^0_{2,3}$, $h^\pm_{1,2}$,
they have only the following three-body decays:
\begin{itemize}
\item For the first family of new fermions:
\bea
N_{1}&\rightarrow&  \nu_e h^0_1,\;  \nu_e Z,\; e^\pm W^{\mp};  \hs
E^\pm_{1}\rightarrow  e^{\pm} h^0_1,\;  e^{\pm} Z,\; \nu_e W^{\mp};
\crn
U_{1}&\rightarrow&  u h^0_1,\;  u Z,\; d W^{+};\hs   D_{1}\rightarrow  d h^0_1,\;  d Z,\; u W^{-},
\label{lnf1decay}
\eea
\item  For the second family of new fermions:
\bea N_{2}&\rightarrow&  \nu_{\mu,\tau} h^0_1,\;  \nu_{\mu, \tau} Z,\; \mu^\pm W^{\mp}\;\tau W^{\mp};\crn
E^\pm_{2}&\rightarrow&  \mu^{\pm} h^0_1,\;\tau^{\pm} h^0_1,\;  \mu^{\pm} Z,\; \tau^{\pm}Z,\; \nu_{\mu,\tau} W^{\mp}; \crn
U_{2}&\rightarrow&   c h^0_1,\;t h^0_1, \;  c Z,\; t Z,\; b W^{+} ,\; s W^{+};\crn
D_{2}&\rightarrow&  b h^0_1,\; s h^0_1,\; b Z,\; s Z,\; c W^{-},\; t W^{-}.
\label{lnf2decay}
\eea
 Because of the suppressed $\Delta_{\tau,b}$, the main  decay modes  are  $F_2\rightarrow f_2h^0_1,f_2W,f_2Z$.
\end{itemize}
The partial decay widths of decays $F\rightarrow fh^0_1, f W, fZ$ are
\bea
\Gamma(F\rightarrow h^0_1f)&=&\frac{m_F}{8\pi}\left|Y_{Ffh^0_1}\right|^2\left(1-\frac{m^2_{h^0_1}}{m^2_F}\right)^2,
\crn
\Gamma(F\rightarrow Vf)&=&\frac{m^3_F}{32\pi m_V^2}\left|g_{FfV}\right|^2\left(1-\frac{m^2_{V}}{m^2_F}\right)^2,
\label{pgaFf}
\eea
where $V=W,Z$,  $ g_{FfV}$ and  $Y_{Ffh^0_1}$ are, respectively, couplings of fermions with gauge and Higgs bosons
given in Tables~\ref{tZff}, \ref{tWff} and \ref{h01SM}. Additional factors 3 are included for quark decays. The decays listed in~(\ref{lnf1decay}) and  (\ref{lnf2decay})
always have $ g_{FfV}\sim \zeta'\epsilon^2$ and $ Y_{Ffh^0_1}\sim s_h \sim \epsilon$,
leading to the consequence that $ \Gamma(F\rightarrow Vf)/\Gamma(F\rightarrow h^0_1f) \sim \zeta'^2\leq \mathcal{O}(10^{-4}) $,
with $\zeta'$ satisfying~(\ref{allowg221}). Hence, every heavy fermion will decay mainly into a light fermion and
a SM-like Higgs boson.

Heavy fermions have been being searched for at the LHC recently, for example the  heavy lepton decays into pairs of light
leptons and the SM-like gauge bosons ~\cite{FeZsearch},  and the null result is consistent with
this investigation. Other heavy quark decays listed in~(\ref{lnf2decay}) are
$U_2\rightarrow h^0_1 t$~\cite{Exuqtoth01}, $ U_2\rightarrow Wb$~\cite{LHCWb}, and $U_2, D_2\rightarrow Zt,Zb$~\cite{CMSQZtb}.
But the promoting channels predicted from this discussion are only $U_2\rightarrow ch^0_1$ and $D_2\rightarrow b h^0_1$.

In conclusion, we have indicated that the allowed regions of parameters given in~\cite{g221} predict
following main fermion decays:
$E_1\rightarrow h^0_1 e$, $U_1,D_1\rightarrow h^0_1u, h^0_1d$, $E_2\rightarrow h^0_1 \mu$,
and $U_2, D_2\rightarrow h^0_1 c,h^0_1 b$. According to our knowledge, these decay channels were not
treated experimentally. We emphasize that this discussion is valid for heavy fermions lighter
than all other new bosons. Any fermions that are heavier than a heavy gauge boson or a Higgs boson
will decay mainly into light fermions and this boson.
\subsection{ Searches for new Higgs bosons at colliders}
At the LHC, the promoting possibility of detecting $h^0_2$ coupling strongly
with heavy fermions  was indicated in~\cite{g221}.  These large couplings are shown in Table~\ref{h023}, where  only large couplings
of neutral CP-even Higgs bosons are shown for  investigating Higgs productions.
\begin{table}[h]
  \centering
  \begin{tabular}{|c|c|}
    \hline
    Vertex & coupling \\
     \hline
   $h^0_2 \overline{F_1}F_1$  & $-c_h c_{\beta'}^2\widetilde{M}_{L_1,Q_1}$ \\
    \hline
    $h^0_2 \overline{F_2}F_2$  & $-c_h c_{\beta'}^2(\Delta^2_{\mu,s}+\Delta^2_{\tau,b})\widetilde{M}_{L_2,Q_2}$ \\
    \hline
     $h^0_2 h^+_1h^-_1$  & $ -c_h\left( \mu c_{\beta}s_{\beta}+\lambda_5c^2_{\beta}u +\lambda_6s^2_{\beta}u \right)$ \\
     \hline
     $h^0_2 h^+_2h^-_2$  & $ -c_h\lambda'_3u$ \\
       \hline
      $h^0_3 \overline{f}f$  & $ \frac{m_f}{vt_{\beta}}$ \\
       \hline
     $h^0_3 h^+_1h^-_1$  & $ -c_{\beta}s_{\beta}\left[\left(\lambda_4-\lambda_1\right)c^2_{\beta} +\left(\lambda_2-\lambda_4\right)s^2_{\beta} \right] v$ \\
        \hline
        $h^0_3 h^+_2h^-_2$  & $ c_{\beta}\left(\lambda'_2-\lambda'_1 \right) v$ \\
        \hline
  \end{tabular}
  \caption{ Possible large couplings of $h^0_{2,3}$ with fermions and charged Higgs bosons}
\label{h023}
\end{table}

We will focus on the remaining new Higgs bosons, including  $h^0_3$, $h_a$, $h^\pm_1$ and $h^\pm_2$. It turns out that they inherit many properties of the new Higgs bosons predicted in THDMs and the MSSM, except $h^\pm_2$. The Higgs sector of the Zee models  can be regarded as the one of a THDM plus a pair of singly charged Higgs bosons, as  investigated thoroughly in Ref.~\cite{chzee}. And the complete investigation of the Higgs phenomenology of the MSSM was presented in~\cite{higgsprodecay} including a brief  comparison with Higgs sector in THDMs. The Higgs bosons $h^0_3$, $h_a$, and $h^\pm_1$ have degenerate masses containing the factor of the trilinear
Higgs self-coupling $\mu$. This property is the same as that of the MSSM, but completely different from THDMs.
Both Refs.~\cite{chzee} and \cite{higgsprodecay} considered the Yukawa part of the THDM type II, where up and down
right-handed singlets of light fermions couple with different Higgs doublets. In contrast in the model under consideration,
all right-handed fermions couple with the same Higgs doublet $\phi$. This explains why the couplings of the neutral $h^0_3$ and $h^\pm_1$ with all  quarks always contain the same factor $\frac{1}{t_{\beta}}$, and  couplings of $h_a$ with all SM-like fermions contain the same
factor $1/_{t_{\beta}}$, as shown in Table~\ref{newhff}. While, the couplings of up and down quarks in the MSSM and THDM type II have different factors of $1/t_{\beta}$ and $t_{\beta}$, respectively.
\begin{table}[h]
  \centering
  \begin{tabular}{|c|c|}
    \hline
   Vertex & Coupling \\
   \hline
    $h_a \overline{e_i}e_i$, $e_i=e,\mu,\tau$  & $\frac{i  m_{e_i} c_{\zeta}}{vt_{\beta}}\left(P_L-P_R\right)$ \\
    \hline
     $h_a \overline{u_i}u_i$,  $h_a \overline{d_i}d_i$, $i=1,2,3$  & $\mp \frac{i  m_{u_i,d_i} c_{\zeta}}{vt_{\beta}}\left(P_L-P_R\right)$ \\
    \hline
   $h_a\overline{L_1}L_1$,  $h_a\overline{Q_1}Q_1$ & $i s_{\zeta}c^2_{\beta'}\widetilde{M}_{L_1,Q_1}(P_L-P_R)$\\
    \hline
      $h_a\overline{L_2}L_2$& $i c^2_{\beta'}(\Delta^2_{\mu}+ \Delta^2_{\tau}) \widetilde{M}_{L_2}s_{\zeta} \left(P_L-P_R\right)$  \\
      \hline
      $h_a\overline{U_2}U_2$, $h_a\overline{D_2}D_2$& $i c^2_{\beta'}\left[(\Delta^2_{s}+ \Delta^2_{b}) \widetilde{M}_{Q_2}s_{\zeta} - \frac{c_{\beta}\left(\Delta^2_{s}m^2_{c,s} +\Delta^2_{b}m^2_{t,b} \right)}{\rho^2_{sb}s^2_{\beta}v^2}\right] \left(P_L-P_R\right)$  \\
       \hline
  $h_a \overline{\nu_e}N_1$, $h_a \overline{e}E_1$ & $ -is_{\zeta}s_{\beta'}c_{\beta'}\widetilde{M}_{L_1}P_L$ \\
  \hline
  $h_a \overline{\mu}E_2$, $h_a \overline{\tau}E_2$ &  $ -ic_{\zeta} c_{\beta'}\Delta_{\mu,\tau}\left[\frac{ m_{\mu,\tau}}{v t_{\beta}\rho_{\mu\tau}}P_R-\rho_{\mu\tau}\widetilde{M}_{L_2}s_{\zeta}P_L -\frac{c_{\beta}c_{\zeta}m^2_{\mu,\tau}}{\rho_{\mu\tau}s^2_{\beta}\widetilde{M}_{L_2}v^2} \epsilon P_L \right]$ \\
   \hline
    $h_a \overline{\nu_\mu}N_2$, $h_a \overline{\nu_\tau}N_2$ &  $ ic_{\zeta} c_{\beta'}\Delta_{\mu,\tau}\rho_{\mu\tau}\widetilde{M}_{L_2}s_{\zeta}P_L$ \\
   \hline
   $h_a \overline{u}U_1$, $h_a \overline{d}D_1$ & $ -is_{\zeta}s_{\beta'}c_{\beta'}\widetilde{M}_{Q_1}P_L$ \\
   \hline
  $h_a \overline{c}U_2$, $h_a \overline{t}U_2$ &  $ -ic_{\zeta} c_{\beta'}\Delta_{s,b}\left[\frac{ m_{c,t}}{v t_{\beta}\rho_{sb}}P_R-\rho_{sb}\widetilde{M}_{Q_2}s_{\zeta}P_L -\frac{c_{\beta}c_{\zeta}m^2_{c,t}}{\rho_{sb}s^2_{\beta}\widetilde{M}_{Q_2}v^2} \epsilon P_L \right]$ \\
  \hline
   $h_a \overline{s}D_2$, $h_a \overline{b}D_2$ &  $ -ic_{\zeta} c_{\beta'}\Delta_{s,b}\left[\frac{ m_{s,b}}{v t_{\beta}\rho_{sb}}P_R-\rho_{sb}\widetilde{M}_{Q_2}s_{\zeta}P_L -\frac{c_{\beta}c_{\zeta}m^2_{s,b}}{\rho_{sb}s^2_{\beta}\widetilde{M}_{Q_2}v^2} \epsilon P_L \right]$ \\
  \hline
  $h_a h^+_{1} W^{\mu}$& $ic_{\xi}s^2_{\zeta}(p_a-p_+)_{\mu}$\\
  \hline
  $h_a h^+_{2} W^{\mu}$& $is_{\xi}s^2_{\zeta}(p_a-p_+)_{\mu}$\\
  \hline
  $h_a h^0_{1} Z^{\mu}$& $\frac{is_{Z}c_h c_{\zeta}c_{\beta}s_{\beta}g(p_a-p_0)_{\mu}}{2s_{\beta'}c_{\beta'}}$\\
  \hline
  \end{tabular}
     \caption{ Couplings of CP-odd  neutral  Higgs bosons, where $p_a,\; p_+$ and $p_0$ are respective incoming momenta of $h_a$, $h^+_{1,2}$, and $h^0_1$.}
\label{newhff}
\end{table}
The notation $\beta$ in this work is equivalent to  $1/t_{\beta}$ defined in ~\cite{chzee,higgsprodecay}, where the  allowed $t_{\beta}$ is consistent with  the constraint (\ref{tzxw}).

Now the recent searches for Higgs bosons in THDMs and MSSM will be used for predictions of detecting
new Higgs bosons discussed in this work. We consider only  Higgs bosons heavier than the top quark. Possible main decays are
\bea
\Gamma(h\rightarrow\overline{ f_1}f_2) &=& \frac{m_h}{8\pi}K(x_1,x_2)\left[1-(\sqrt{x_1}+\sqrt{x_2})^2\right]
\times \left|Y_{hff}\right|^2,  \crn
\Gamma(h\rightarrow V_1V_2)&=& \frac{m_{h}}{8 \pi } K(x_1,x_2) \left[ 8x_1x_2+(1-x_1-x_2)^2 \right]
\times \frac{m^2_{h}\left|g_{hVV}\right|^2}{8m^2_{V_1}m^2_{V_2}},\crn
\Gamma(h_1\rightarrow  h_2 V)&=& \frac{m_{h_1}}{8 \pi } K(x_1,x_2) \left[ (1-x_1)^2 -x_2(2+2x_1-x_2)\right]
\times \frac{m^2_{h_1}\left|g_{hhV}\right|^2}{2m^2_{V}}, \crn
\Gamma(h\rightarrow  h_1 h_2)&=& \frac{m_h}{8 \pi}K(x_1,x_2)\times  \left|\frac{\lambda_{hhh}}{ m_{h}}\right|^2 ,
\label{hdecay}
\eea
where $K(x_1,x_2)=\left[(1-x_1-x_2)^2 -4 x_1x_2\right]^{\frac{1}{2}}$; $x_{1,2}=\frac{m_{1,2}^2}{m_0^2}$;
$m_{0}$ and $m_{1,2}$ denote the masses of the initial and final states, respectively.  The factor $1/2$
is applied if the two final state particles are identical.  A final massless state gives $K(x_1,0)=(1-x_1)^{\frac{1}{2}}$.
Expressions for couplings  Yukawa
 $Y_{hff}$,
   gauge-Higgs-Higgs  $g_{hVV}$,  the Higgs-Higgs-gauge  $g_{hhV}$, and  $\lambda_{hhh}$ were listed in the above Tables.   The correlations between the different partial decay widths
of a Higgs boson depend only on the last factors of formulas in~(\ref{hdecay}). Hence, they will be used to
estimate the largest partial decay widths.

The main decay channels of $h^\pm_1$ are $h^+_1\rightarrow \overline{t}b, Z'W,ZW'$ have  relative factors as follows:
\[
\left|Y_{h^\pm_1 tb}\right|^2=\left|\frac{\sqrt{2m_t}}{vt_{\beta}}\right|^2\simeq \frac{1}{t^2_{\beta}}, \hs
\frac{m^2_{h^\pm_1}\left|g_{h^\pm_1ZW'}\right|^2}{8m^2_{Z}m^2_{W'}};\;
\frac{m^2_{h^\pm_1}\left|g_{h^\pm_1Z'W}\right|^2}{8m^2_{Z'}m^2_{W}} \simeq \frac{g^2}{8} \times \frac{m^2_{h^\pm_1}}{m^2_{W'}},
\]
where the allowed values of $t_{\beta}$ are given in~(\ref{tzxw}). Hence, if $m_{h^\pm_1}$ is not too larger than
the  heavy gauge boson masses, the main decay  is $h^+_1\rightarrow \overline{t}b$,
where the $h^\pm_1 tb$ coupling is the same as in the MSSM.
The LHC has searched for this decay recently~\cite{exchtotb,exchtotb2}, through the production channel
$pp\rightarrow tbh^\pm$, giving the lower bound of 1~TeV for  $m_{h^\pm_1}$.
\section{Conclusion}
\label{conclusion}
Recently, the G221 model has been introduced in Ref.~\cite{g221}  with the main purpose to explain all
experimental data in flavor physics, tau decays, electroweak precision data, and LNU phenomenology from
the anomalies in $B$ decays. But there are still to crucial questions to this model, namely, how to generate
active neutrino masses and DM? This work indicated that these problems can be solved based on the mechanisms
of generating the active neutrino masses by radiative corrections. In particular, the simplest way to generate
the active neutrino masses based on the Zee models was shown in detail. The model predicts the existence of
a new pair of singly charged Higgs bosons that have large couplings only with light leptons and Higgs bosons.
The DM problem can be solved by applying similar mechanisms shown in many radiative neutrino mass models
with DM that were widely investigated previously.

In this work we have analyzed a more general diagonalization of gauge boson mass matrices.
We have found that the ratio of the tangents of  $Z-Z'$ and $W-W'$ mixing angles
is the cosine of the Weinberg angle, $ \cos \theta_W$.
This leads to the consequence that the number of the model parameters is reduced by 1.

The most important results of this work were obtained in the Higgs sector where new interesting properties
of physical Higgs bosons were explored. First, using the minimal conditions of the Higgs potential to cancel
all mutually dependent parameters in the potential, we found that the two squared mass matrices of singly charged
and neutral CP-odd Higgs bosons are proportional to the coefficient of the triple Higgs couplings $\mu$. Second, the masses and  physical states of all Higgs bosons,
as well as their mixing matrices, were presented clearly so that all couplings of the Higgs bosons with the remaining
particles can be determined. From this, the SM-like Higgs boson and its couplings were easily identified and
compared with experimental data, leading to the important constraint on the mixing parameter $c_h$, namely $0.995<|c_h|<1$.
Regarding the new Higgs bosons, three Higgs bosons $h^0_1$, $h_a$ and $h^\pm_1$ have degenerate masses.
Namely, the analytic expression for the squared mass is $ \frac{\mu u }{s_{2\beta}}+ v^2 s_{\beta}c_{\beta}$,
where the main contribution has the same form as the new Higgs boson masses in the MSSM.
In addition, their coupling properties are the same as in the THDM of type I.
Hence, their behaviors can be predicted based on well-known studies of the THDM as well as of the MSSM.

We combined the above results and the allowed regions of parameters indicated in Ref.~\cite{g221}
to predict some promoting decay channels of new fermions and Higgs bosons. We found that the decays
of new heavy particles to SM-like gauge bosons are very suppressed, due to the very small mixing of
heavy and SM gauge bosons. The main decays of heavy fermions into two SM-like particles are the decays
$F_{1,2}\rightarrow h^0_{1} f_{1,2}$. Decays into SM-like fermions in the third family are very suppressed
because the allowed regions contain the tiny coefficient $\Delta_{\tau,b}$.  The main decay of $h^\pm_1$ is
$h^\pm\rightarrow tb$. The latest searches for this decay channel give a  $1$~TeV lower bound
for the charged Higgs boson mass.

The LHC have searched for many decay channels of new fermions into SM-like fermions of the third family.
So the model will be checked by experiments in coming years. If these decay channels are detected, the model must be extended.
For example, the third family of new vector-like fermions should be added to release the allowed regions of parameters.
\section*{Acknowledgments}
H. N. L. thanks BLTP, JINR for financial support and hospitality during his stay where this work is performed.
 This research is funded by Vietnam  National Foundation for Science and Technology Development (NAFOSTED)
under Grant number  103.01-2017.22.
%
\appendix
\section{\label{VLij} Masses and mixing parameters of charged leptons}
If $\lambda_{\ell}$ has the form given in Eq. (\ref{laell1}), the precise formula of $V^{11}_L$ defined in (\ref{VL}) is
\be
V^{11}_L=\sqrt{I_3-\frac{1}{4}\la_{\ell}\widetilde{ M}^{-2}_L\la^{\dagger}_{\ell}}= \left(
                \begin{array}{ccc}
                   s_{\beta'} & 0 & 0 \\
                  0 & 1-\frac{c^2_{\beta'}\Delta^2_{\mu}}{1+ \rho_{\mu\tau}} &-\frac{c^2_{\beta'}\Delta_{\mu}\Delta_{\tau}}{1+\rho_{\mu\tau}} \\
                  0 & -\frac{c^2_{\beta'}\Delta_{\mu}\Delta_{\tau}}{1+\rho_{\mu\tau}} &  1-\frac{c^2_{\beta'}\Delta^2_{\tau}}{1+\rho_{\mu\tau}} \\
                \end{array}
              \right),
\label{V11L1}
\ee
where $\rho_{\mu\tau}=\sqrt{1-c^2_{\beta'}\left(\Delta^2_{\mu}+\Delta^2_{\tau}\right)}$. Other submatrices contained in $V_L$ are

\bea
V^{12}_L&=&\left(
                 \begin{array}{cc}
                    -\frac{s^2_{\beta}\widetilde{M}_{L_1}u}{M_{L_1}}& 0 \\
                   0 &-\frac{\Delta_{\mu}s_{\beta'}\widetilde{M}_{L_2}u\rho_{\mu\tau}}{M_{L_2}}  \\
                   0 &  -\frac{\Delta_{\tau}s_{\beta'}\widetilde{M}_{L_2}u\rho_{\mu\tau}}{M_{L_2}}\\
                 \end{array}
               \right), \hs
 V^{21}_L  =\left(
                \begin{array}{ccc}
                  s_{\beta'} & 0 & 0 \\
                  0 & s_{\beta'}\Delta_{\mu} & s_{\beta'}\Delta_{\tau} \\
                \end{array}
              \right),
 \crn
V^{22}_L &=&\mathrm{diag}\left( \frac{M_{L_1}}{\widetilde{M}_{L_1}u},\; \frac{M_{L_2}}{\widetilde{M}_{L_2}u}\right).\nn
\eea
After the block-diagonalization, the SM blocks  of fermions matrices must satisfy the experimental constraints. In general, the SM block of the charged lepton mass matrix  $ V_eV_L\mathcal{M}_{\mathcal{E}}W^{\dagger}_e=\mathcal{M}'_{\mathcal{E}}$
will not be  diagonal  if the matrix $y_{\ell}$ in (\ref{g221e57}) is assumed to be diagonal for simplicity.  Instead of,  $y_{\ell}$ is chosen so that only mixing on $\mu-\tau$ sector is non-zero, the corresponding SM block of   $\mathcal{M}'_{\mathcal{E}}$ is
\be
M_{\ell} \simeq \frac{v s_{\beta}}{\sqrt{2}}\left(
                \begin{array}{ccc}
                  y_{e}s_{\beta'} &0  & 0 \\
                 0  & y_{\mu}\left[ 1-\frac{c^2_{\beta'}\Delta_{\mu}(\Delta_{\mu}+ \Delta_{\tau}y_{\tau\mu}/y_{\mu})}{1+\rho_{\mu\tau}}\right] & y_{\mu\tau}-\frac{c^2_{\beta'}\Delta_{\mu} (\Delta_{\mu}y_{\mu\tau} + \Delta_{\tau}y_{\mu})}{1+\rho_{\mu\tau}} \\
                 0  & y_{\tau\mu}-\frac{c^2_{\beta'}\Delta_{\tau} (\Delta_{\mu}y_{\tau\mu} + \Delta_{\tau}y_{\tau})}{1+\rho_{\mu\tau}} &y_{\tau}\left[ 1-\frac{c^2_{\beta'}\Delta_{\tau}(\Delta_{\tau}+ \Delta_{\mu}y_{\mu\tau}/y_{\tau})}{1+\rho_{\mu\tau}}\right]  \\
                \end{array}
              \right).
  \label{llepm}
\ee
There exist values of  $y_{\mu\tau,\tau\mu}$ so that the matrix~(\ref{llepm}) is  diagonal and  the result of~\cite{g221} is unchanged.
The diagonal SM block of charged leptons also guarantees that the lepton flavor violating decay $h^0_1\rightarrow\mu\tau$ is  suppressed, consistent with  experimental constraints.
Then $y_{\mu\tau}$ and $y_{\tau\mu}$ are chosen to satisfy the condition $(M_{\ell})_{23}=(M_{\ell})_{32}=0$.
Now the elements of the Yukawa coupling matrix $y_\ell$ can be expressed as
\bea
y_{e}&=&\frac{\sqrt{2}m_e}{vs_{\beta}s_{\beta'}},  \;
 y_{\mu, \tau}=\frac{\sqrt{2}m_{\mu,\tau}}{vs_{\beta}} \times \frac{1}{\Delta^2_{\tau}+\Delta^2_{\mu}} \left[\Delta^2_{\tau,\mu}
+ \frac{\Delta^2_{\mu,\tau}}{\rho_{\mu\tau}}\right],
\crn
y_{\mu\tau,\tau\mu} &=&- \frac{\sqrt{2}m_{\tau,\mu}}{vs_{\beta}} \times \frac{\Delta_{\mu}\Delta_{\tau}}{\Delta^2_{\tau}+\Delta^2_{\mu}}
\left[1-  \frac{1}{\rho_{\mu\tau}}\right].
\label{yell2}
\eea
\section{\label{zeem}Neutrino masses from one-loop corrections }
First, we consider the simplest case where  $\varphi^\pm$, $\delta^\pm$ and charged leptons in Fig.~\ref{acNuMass}
are all mass eigenstates; the one-loop amplitude contributing to the neutrino masses is
\bea
i \frac{1}{2}(m_{\nu})_{ba}\overline{\nu_{L_b}} (\nu_{L_a})^c &\equiv& \int\frac{d^4p}{(2\pi)^4}\times
\overline{\nu_{L_b}} [-i(y_{\ell})_{db}] \frac{im_{e_d}}{p^2-m^2_{e_d}} [-i2f_{da}](\nu_{L_a})^c
\crn
&&\times\frac{i}{p^2-m^2_{\delta}}\left(-i\lambda_{\delta}\frac{uvc_{\beta}}{2}\right)\frac{i}{p^2-m^2_{\varphi^\pm}},
\label{mnumass}
\eea
where we have used  $\langle\Phi^0\rangle=\frac{u}{\sqrt{2}}$,
$\langle \varphi'^0\rangle=\frac{ v c_{\beta}}{\sqrt{2}}$ and masses of the charged leptons
$m_{e_d}=(y_{\ell})_{cd}\langle\varphi^0\rangle$. The right hand side of (\ref{mnumass}) is rewritten as
\bea
i(m'_{\nu})_{ba} &\equiv& 2m_{e_d}f_{da} \lambda_{\delta}c_{\beta}(y_\ell)_{db}uv\int\frac{d^4p}{(2\pi)^4}
\frac{1}{\left(p^2-m^2_{\delta}\right)\left(p^2-m^2_{\varphi^\pm}\right)\left(p^2-m^2_{e_d}\right)}
\crn
&&= 2m_{e_d}f_{da} \lambda_{\delta}c_{\beta}(y_\ell)_{db} \frac{uv}{16\pi^2}\times
\frac{1}{m^2_{\varphi^\pm}-m^2_{\delta}} \ln\left[\frac{m^2_{\delta}}{m^2_{\varphi^\pm}}\right],  \nn
\label{mnumass1}
\eea
where
\be
\left. \int\frac{d^4p}{(2\pi)^4} \frac{1}{\left(p^2-m^2_{\delta}\right)\left(p^2-m^2_{\varphi^\pm}\right)
\left(p^2-m^2_{e_d}\right)}\right|_{m^2_{e_d}\rightarrow0}=\frac{i}{16\pi^2}\times
\frac{1}{m^2_{\varphi^\pm}-m^2_{\delta}} \ln\left[\frac{m^2_{\delta}}{m^2_{\varphi^\pm}}\right].
\label{c0}
\ee
Because $\overline{\nu_{L_b}} (\nu_{L_a})^c= \overline{\nu_{L_a}} (\nu_{L_b})^c$, the mass matrix $(m_{\nu})_{ab}$
is written in the symmetric form as follows: $(m_{\nu})_{ab}=(m_{\nu})_{ba}=\frac{1}{2}\left[(m'_{\nu})_{ab}+(m'_{\nu})_{ba}\right]$.
In the simple case, where $ (y_{\ell})_{db}=\delta_{db}m_{e_b}/\langle \varphi^0\rangle=\delta_{db}m_{e_b}\sqrt{2}/(vs_{\beta})$,
we can use the antisymmetric property  $f_{ad}=-f_{da}$ to write the neutrino mass matrix in the following form:
\be
(m_{\nu})_{ba} = \frac{f_{ba}\lambda_{\delta}\sqrt{2}}{16\pi^2t_{\beta}}\times
\frac{u \left(m^2_{e_b}- m^2_{e_a}\right)}{m^2_{\varphi^\pm}-m^2_{\delta}} \ln\left[\frac{m^2_{\delta}}{m^2_{\varphi^\pm}}\right].
\label{mnu2}
\ee
If  $m_{\delta}=m_{\varphi^\pm}$, then  $\lim_{m_{\delta}\rightarrow m_{\varphi^\pm}} \frac{1}{m^2_{\varphi^\pm}- m^2_{\delta}} \ln\left[\frac{m^2_{\delta}}{m^2_{\varphi^\pm}}\right]
= -\frac{1}{m^2_{\varphi^\pm}},$
leading to a simpler form of (\ref{mnu2}).

\end{document}